\documentclass[oneside,11pt]{article}
\usepackage{epsfig}
\begin{document}

\begin{center}
\begin{LARGE}
Coherent Beam-Beam Tune Shift of Unsymmetrical Beam-Beam 
Interactions with Large Beam-Beam Parameter
\end{LARGE}

\vspace{0.25in}
Lihui Jin and Jicong Shi

Department of Physics \& Astronomy, The University of Kansas, 
Lawrence, KS 66045

Georg H. Hoffstaetter 

Department of Physics, Cornell University, Ithaca, NY 14850
\end{center}

\begin{center} 
Abstract 
\end{center}

Coherent beam-beam tune shift of unsymmetrical beam-beam interactions was 
studied experimentally and numerically in HERA where the lepton beam has a 
very large beam-beam parameter (up to $\xi_y=0.272$). Unlike the symmetrical 
case of beam-beam interactions, the ratio of the coherent and incoherent 
beam-beam tune shift in this unsymmetrical case of beam-beam interactions was 
found to decrease monotonically with increase of the beam-beam parameter. The 
results of self-consistent beam-beam simulation, the linearized Vlasov equation, 
and the rigid-beam model were compared with the experimental measurement. It 
was found that the coherent beam-beam tune shifts measured in the experiment 
and calculated in the simulation agree remarkably well but they are much smaller 
than those calculated by the linearized Vlasov equation with the single-mode 
approximation or the rigid-beam model. The study indicated that the single-mode 
approximation in the linearization of Vlasov equation is not valid in the case 
of unsymmetrical beam-beam interactions. The rigid-beam model is valid only 
with a small beam-beam parameter in the case of unsymmetrical beam-beam 
interactions.

\vspace{0.25in}

\vspace{0.1in}
\noindent
Corresponding author:

Jack Shi,  
Department of Physics \& Astronomy, University of Kansas, Lawrence, KS 66045

\hspace*{0.6in} Phone: 785-864-5273; Fax: 785-864-5262; Email: jshi@ku.edu

\newpage

\begin{center}
{\bf I. INTRODUCTION}
\end{center}

To achieve a substantial increase of luminosity in a storage-ring collider, 
limited options include increase of bunch currents, reduction of beam sizes 
at interaction points (IPs), and increase of the number of colliding bunches.
The first two measures unavoidably increase head-on beam-beam forces 
which could lead to collective (coherent) beam-beam instabilities 
\cite{Chao,Shi}. Understanding of coherent beam-beam effects especially in 
the nonlinear regime is therefore of primary importance for achieving high 
luminosity in a storage-ring collider with high-intensity beams. 

To study the coherent beam-beam effect, one important quantity that can be 
measured experimentally is the coherent beam-beam tune shift. Without 
beam-beam interactions and without considering nonlinearities in the lattice, 
the two counter-rotating beams oscillate transversely with frequencies that 
correspond to lattice tunes (betatron tunes without collision) if they 
deviate from close orbits. With beam-beam interactions, the particle 
distributions of the beams are perturbed and evolve with time according to 
the Vlasov equation \cite{Chao}. The dynamics of the beams could therefore 
be complicated by multi-mode oscillations of the beam distributions. If 
considering only the stable oscillation of beam centroids (coherent dipole 
oscillation), the frequency spectrum of the beam-centroid oscillation has 
two primary frequencies for each degree of freedom of the transverse motion. 
These primary frequencies correspond to the tunes measured during collision. 
The differences between these measured collision tunes and the lattice tunes 
are the coherent beam-beam tune shifts \cite{Chao,Meller,Yokoya}. Over decades, 
many studies have been conducted on the relationship between the coherent 
beam-beam tune shift and the beam-beam parameter that measures the strength of 
the beam-beam interaction [3-13]. Two theoretical models, the linearized Vlasov 
equation \cite{Chao} and the rigid-beam model \cite{Hirata}, have been studied 
extensively for cases of weak beam-beam perturbation in which the beam-beam 
parameter is relatively small. When the two beams have the same or very close 
lattice tunes, the calculation of the coherent beam-beam tune shift based on 
the linearized Vlasov equation with single-mode approximation agrees with 
beam measurements and computer simulations \cite{Meller,Yokoya,KEK,Keil}. The 
rigid-beam model is inconsistent with the linearized Vlasov equation and was 
therefore proven to be wrong by beam measurements in this case \cite{KEK,Keil}.
When the two beams have very different lattice tunes, on the other hand, the
calculation based on the rigid-beam model provides a good agreement with beam
measurements \cite{Hirata,Hofmann,Georg99}. In both of these models, the 
equilibrium beam distributions were assumed to be Gaussian distributions for 
easing the calculations. In the case of weak beam-beam perturbation, this 
assumption is fairly good as the beams were observed to stay close to a 
Gaussian. 

The situation of strong beam-beam perturbations with relatively large beam-beam 
parameter is much more complicated and less understood. When the beam-beam 
parameter exceeds a threshold, the beam-beam interaction could induce a chaotic 
coherent beam-beam instability. After the onset of the instability, the closed 
orbits could become unstable for the beam centroids and two beams could develop 
a spontaneous unstable coherent oscillation \cite{Shi,Jin}. When the beam-beam 
parameter is below the beam-beam threshold, the coherent beam oscillation is 
stable. It is, however, not clear whether the linearized Vlasov equation or 
the rigid-beam model are still valid in the regime of strong beam-beam 
perturbation. As many efforts are being made to further increase the beam-beam 
parameter in upgrades of current and developments of future storage-ring 
colliders, an understanding of the coherent beam-beam tune shift in this 
regime not only is necessary for the interpretation of the tune measurement 
during operation of colliders with high-intensity beams but also could shed 
light on the onset of the chaotic coherent beam-beam stability.

To explore the beam-beam effect with a large beam-beam parameter a beam
experiment, the HERA 2000 beam study, was performed on HERA [Hadron Electron 
Ring Accelerator at DESY (Deutsches Elektronen-Synchrotron), Hamburg, Germany] 
in which a 920 GeV proton ($p$) beam and a 27.5 GeV positron ($e^+$) beam 
collided at two IPs, H1 and ZEUS \cite{Georg00}. The beam-beam interaction in 
HERA is a typical case of unsymmetrical beam-beam interaction as the two beams 
have very different lattice tunes and beam-beam parameters (strong $p$ beam 
and weak $e^+$ beam). In the experiment, the vertical beam-beam parameter of 
the $e^+$ beam was varied from 0.068 to 0.272 by changing the vertical 
beta-function of the $e^+$ beam at two IPs. The emittance of the $e^+$ beam 
and the luminosity were measured as functions of the beam-beam parameter. 
One important phenomenon observed in this experiment is that the measured 
coherent beam-beam tune shifts of the $e^+$ beam are much smaller than those 
calculated from the rigid-beam model. This is the first experimental evidence 
indicating that the traditional models of the coherent beam-beam tune shift 
are no longer valid in the situation of strong beam-beam perturbations. 

To have a better understanding of the experimental data, we reconstructed 
the HERA beam experiment with a self-consistent beam-beam simulation. 
Remarkable agreement between the experiment and the simulation was observed 
on emittance growth and luminosity reduction. More significantly, the computer 
simulation confirmed the experimental result of very small coherent beam-beam
tune shifts in this case of a very large beam-beam parameter. To examine the 
validity of the theoretical models for the coherent beam-beam tune shift, 
the linearized Vlasov equation and the rigid-beam model were solved for the 
HERA experiment. Since the distribution of the $e^+$ beam significantly
deviated from the Gaussian due to the strong beam-beam interaction, 
the solutions of the linearized Vlasov equation and the rigid-beam model 
were calculated with the beam distributions obtained from the beam-beam 
simulation instead of assuming Gaussian distributions. It was found that for 
the unsymmetrical beam-beam interaction with a large beam-beam parameter, 
the result of the rigid-beam model is inconsistent with the beam experiment 
and beam simulation even though a more accurate beam distribution was used 
in the calculation. The coherent beam-beam tune shifts calculated from the 
linearized Vlasov equation with the single-mode approximation were also 
found to be significantly different from the result of the beam experiment 
and beam simulation no mater whether the beam-beam parameter is large or 
small. As the linearization of the Vlasov equation is expected to be valid 
for at least a small beam-beam parameter, this discrepancy suggests that the 
single-mode approximation used for solving the linearized Vlasov equation 
may not be valid in the case of unsymmetrical beam-beam interactions. 
Unfortunately, without the single-mode approximation the linearized Vlasov 
equation for the problem of beam-beam interactions is currently unsolvable
computationally due to the unsolved degeneracy problem of a matrix with mode 
coupling. 

This paper is organized as follows. Section II summarizes the results of 
the HERA 2000 beam experiment. In Section III, the self-consistent beam 
simulation for the HERA beam experiment is discussed. In Section IV, the 
coherent beam-beam tune shifts calculated by using the rigid-beam model or 
the linearized Vlasov equation are compared with the experiment/simulation 
results. The details of the coherent tune calculation with the theoretical 
models are presented in Appendix A and B. The characteristics of the coherent 
beam-beam tune shift in the unsymmetrical case of beam-beam interactions are
discussed in Section V. Section VI contains a summary remark.

\begin{center}
{\bf II. HERA 2000 BEAM EXPERIMENT}
\end{center}

In the luminosity upgrade of HERA, the beam-beam parameters of the electron 
beam have been nearly doubled. To examine any possible luminosity reduction 
due to beam-beam effects, a series of beam experiments were performed in 
HERA \cite{Georg99,Georg00}. In the HERA 2000 beam experiment, the $e^+$ 
beam was used to collide with the $p$ beam at the two IPs and the effect of 
a large beam-beam parameter of the lepton beam was explored by increasing 
the vertical beta-function ($\beta_{e,y}$) of the $e^+$ beam at the IPs. 
The vertical beam-beam parameter ($\xi_{e,y}$) of the $e^+$ beam is related 
to $\beta_{e,y}$ by \cite{Sands}
\begin{equation}
\xi_{e,y} = \frac {r_e N_p}{2\pi\gamma_e}\frac {\beta_{e,y}}{\sigma_{p,y}
(\sigma_{p,x} + \sigma_{p,y})}
\end{equation}
where $r_e$ and $\gamma_e$ are the classic radius and Lorentz factor of
positron, respectively, and $N_p$ is the number of protons per bunch.
The horizontal and vertical size of the $p$ beam at the IPs are given by
$\sigma_{p,x}$ and $\sigma_{p,y}$. In the experiment, the $p$ beam current 
($I_p$) was fixed. Since the beam-beam parameter of the $p$ beam is very small, 
there was little change in the $p$ beam size as it was observed during the 
experiment. The vertical beam-beam parameter $\xi_{e,y}$ is therefore linearly 
proportional to $\beta_{e,y}$ in this case. During the experiment, after the 
proton current was filled, $\xi_{e,y}$ was increased from 0.068 to 0.272 as 
$\beta_{e,y}$ was changed from 1.0 to 4.0 m while other lattice parameters 
were kept as constants. It should be noted that with two IPs in the HERA 
experiment a beam-beam parameter of 0.272 is among the highest ever achieved 
in storage-ring colliders. Table 1 lists the $e^+$ beam current ($I_e$) and 
the beam-beam parameters of the $e^+$ and $p$ beam at $\beta_{e,y}$ where the 
measurement was performed. For the $p$ beam, $I_p=90$ mA, the beta-function 
at the IPs were $(\beta_{p,x},\beta_{p,y})=(7.0,\;0.5)$ m; the lattice 
tunes were $(\nu_x,\nu_y)=(31.291,\;32.297)$; and the emittance without 
collision was $(\epsilon_{0x},\epsilon_{0y})=(3.82,\;3.18)$ nm. For 
the $e^+$ beam, $\beta_{e,x}=2.5$ m, $(\nu_x,\nu_y)=(52.169,\;52.246)$, and 
$(\epsilon_{0x},\epsilon_{0y})=(32.0,\; 1.28)$ nm. Other beam parameters can 
be found in \cite{Georg00}. The $e^+$ and $p$ beam sizes were not matched 
during the collision. In the experiment, the emittance of the $e^+$ beam and 
the luminosity were measured as functions of $\beta_{e,y}$ at both the 
IPs. In Figs. 1 and 2, the measured emittance and the specific luminosity 
were plotted, with discrete points, as functions of $\xi_{e,y}$. For each 
$\xi_{e,y}$ where the measurement was performed, two data points correspond 
to the measurement at the two IPs, respectively. The specific luminosity 
is defined as ${\cal L}_s=N_{col}{\cal L}/(I_eI_p)$ where $N_{col}$ and 
${\cal L}$ are the number of colliding bunches and the luminosity, 
respectively \cite{Georg99}. As shown in Fig. 1, the vertical emittance 
growth of the $e^+$ beam increases monotonically and smoothly with the 
increase of $\xi_{e,y}$. This is the characteristics of the incoherent 
beam-beam effect in contrast to the coherent beam-beam effect of which the 
emittance growth as a function of the beam-beam parameter could experience 
certain jumps (phase transitions) due to the onset of the coherent beam-beam 
instability \cite{Shi}. To confirm that the luminosity reduction in Fig. 2 
is indeed due to the emittance blowup, the luminosity calculated with the 
measured emittance by using the standard formula \cite{Georg00} is also 
plotted in the Fig. 2. The agreement between the measured luminosity and 
the calculated luminosity in Fig. 2 shows a consistency in the emittance 
and luminosity measurement. 

In the experiment, the collision tunes of the $e^+$ beam were also measured 
at $\beta_{e,y}=4.0$ m as $\nu_x=52.160$ and $\nu_y=52.233$. The coherent 
beam-beam tune shift of the $e^+$ beam is therefore only $\Delta\nu_x=0.009$ 
and $\Delta\nu_y=0.013$ while from the rigid-beam model $\Delta\nu_x=0.016$ 
and $\Delta\nu_y=0.042$ if both the beams are Gaussian \cite{Georg99}. The 
measured coherent beam-beam tune shifts in this case are inconsistent with 
the traditional understanding of beam coherent oscillation. Moreover, in the 
symmetrical case of beam-beam interactions, the ratio of the coherent and 
incoherent beam-beam tune shift has a value approximately ranging from 1.2 
for round beams to 1.3 for flat (one-dimensional) beams \cite{Yokoya}. Since 
the beam-beam parameters of the $p$ beam is very small, the distribution 
of the $p$ beam in the HERA experiment is still very close to a Gaussian. 
The incoherent beam-beam tune shifts that are the tune shifts of a single 
particle with an infinitesimal betatron oscillation are $2\xi_{e,x}=0.082$ 
and $2\xi_{e,y}=0.544$ for the $e^+$ beam in the horizontal and vertical 
plane, respectively. The ratios of coherent and incoherent beam-beam tune 
shifts of the $e^+$ beam are $\Delta\nu_{e,x}/(2\xi_{e,x})=0.11$ and 
$\Delta \nu_{e,y}/(2\xi_{e,y})=0.024$ at $\beta_{e,y}=4.0$ m. The coherent 
beam-beam tune shifts in this case of unsymmetrical beam-beam interactions 
with a large beam-beam parameter are therefore extremely small as compared 
with the incoherent beam-beam tune shifts or compared with the symmetrical 
case of beam-beam interactions. 

\begin{center}
{\bf III. RECONSTRUCTION OF HERA BEAM EXPERIMENT WITH NUMERICAL SIMULATION}
\end{center}

To have a better understanding of the measured data in the HERA experiment, we 
have reconstructed the experiment with a self-consistent beam-beam simulation.
This numerical study also served as a detailed benchmark of our beam-beam 
simulation code with the experimental measurement. In the simulation, the 
linear HERA lattice with the two IPs was used. The code used in this study 
is an expanded version of \cite{Shi} that is currently capable of studying 
beam-beam effects of proton or lepton beams with any aspect ratio (ratio 
between vertical and horizontal beam size). The two colliding beams were 
represented by a million macro-particles with given initial Gaussian 
distributions in transverse phase space. Without beam-beam interactions, the 
initial beam distribution used in the simulation matches exactly with the 
lattice. Beam-beam interaction at each the IP was represented by a kick in 
transverse phase space and the kick was calculated by using the 
particle-in-cell method as described in Ref. \cite{Shi}. Since the beams in 
the HERA experiment were flat, a uniform mesh extending to $\pm 20\sigma$ 
in the configuration space with a grid constant of $0.2\sigma$ was necessary 
in this case. Tracking of particle motion was conducted in four-dimensional 
transverse phase space without synchrotron oscillations and momentum 
deviations. For lepton beams, the quantum excitation and synchrotron damping 
were treated as kicks in each turn during the tracking. The horizontal kick 
is \cite{Damping}
\begin{eqnarray}
 \Delta x  & = & e^{-1/(2\tau_x)}x +\left [\left(1-e^{-1/(2\tau_x)}\right)
             \epsilon_x \right ]^{1/2} w_1 \nonumber \\
 \Delta p_x & = & e^{-1/(2\tau_x)}p_x + \left [\left(1-e^{-1/(2\tau_x)}\right)
             \epsilon_x \right ]^{1/2} w_2   
\end{eqnarray}
where $x$ and $p_x$ are the normalized horizontal coordinate and its conjugate 
momentum, $\epsilon_x$ is the horizontal emittance, and $w_1$ and $w_2$ are
random numbers with a Gaussian distribution that is centered at zero and has 
unit standard deviation. The damping time in horizontal and vertical direction,
$\tau_x$ and $\tau_y$, have the unit of turns. For HERA, $\tau_x=436$ and 
$\tau_y=600$, respectively. The vertical kick has a similar formula. 

With the beam-beam simulation, the emittance growth of the $e^+$ beam and the 
specific luminosity were calculated as functions of $\xi_{e,y}$ for the HERA
experiment and the results were plotted in Figs. 1 and 2 as solid lines. Both 
the emittance and the luminosity plot show a remarkable agreement between 
the experiment and the simulation. Figure 3 plots the evolution of the 
vertical emittance of the $e^+$ beam at different $\beta_{e,y}$. In all these 
cases, after a quick emittance blowup, the beam emittance is re-stabilized 
and, consequently, an equilibrium (or quasi-equilibrium) state of the $e^+$ 
beam was reached. During the emittance blowup, the particles in the beam core
escape quickly to the beam tails. Without the onset of the coherent beam-beam 
instability, the particles in the beam tails are stable for the beam-beam 
interaction. This re-stabilization of the beam emittance is therefore due 
to a depopulation of the beam core and is consistent with the experimental
observation that the beam lifetime and operation conditions were good during
the experiment even in the case of $\xi_{e,y}=0.272$ \cite{Georg00}. 

In the beam simulation, the coherent beam-beam tune shift was also calculated.
Fig. 4 is the calculated power spectrum of the coherent oscillation of the
$e^+$ beam at $\beta_{e,y}=4.0$ m. Due to the quantum fluctuation, the $e^+$ 
beam has always a very small oscillation which is enough for the calculation 
of the coherent frequency if the numerical noise is small. Since a large 
number of particles were used for each of the beams, in this simulation 
the numerical noise was very low and we were able to calculate the coherent
frequency without applying off-center kicks on the beams. The spectrum
was calculated by the fast Fourier transformation (FFT) of the beam centroid 
motion from the 5000th to 9000th turn. Since at the 5000th turn the beam has
already reached its quasi-equilibrium, the transient state of the beam was 
thrown out during the tune calculation (see Fig. 3). As shown in Fig. 4 the 
power spectrum peaks at 0.1605 and 0.2331 in the horizontal and vertical 
planes, respectively, which corresponds to $\nu_x=52.161$ and 
$\nu_y=52.233$ for the beam coherent tunes during collision. This simulation
result agrees excellently with the experimental measurement of $\nu_x=52.160$ 
and $\nu_y=52.233$. The beam-beam simulation therefore confirmed the coherent 
beam-beam tune shift measured in the HERA beam experiment. 

To understand the very small coherent beam-beam tune shift in this case of 
very large beam-beam parameter, the dynamics of the beam particle 
distributions was studied during the beam simulation. As shown in Fig. 5, 
the distribution of the $e^+$ beam deviates from a Gaussian distribution with 
a significant drop at the beam core and a growth of the beam tails. Compared 
with the distribution of the $e^+$ beam, a Gaussian beam that has the same 
emittance of the $e^+$ beam has more particles in the core. The real coherent 
beam-beam tune shift measured in the experiment and calculated with the beam
simulation is therefore smaller than that calculated from the rigid-beam 
model with Gaussian beams. Moreover, the assumption of Gaussian beams in all
the theoretical models of the coherent beam-beam oscillation neglects the
effect of the $\phi$-dependence of the beam equilibrium distributions, where
$\phi$ is the angle of the action-angle variable. In the HERA beam experiment, 
the equilibrium (quasi-equilibrium) distribution of the $e^+$ beam in fact 
has a strong $\phi$-dependence as shown in Fig. 6. For a comparison, the 
initial distribution used in the simulation was also plotted in the Figure 
and shows no $\phi$-dependence as it should be. In order to compare 
the experimental/simulation result with calculations of theoretical models, 
the theoretical models need to be modified to include the $\phi$-dependence 
of non-Gaussian equilibrium distributions.

\begin{center}
{\bf IV. VALIDITY OF THEORETICAL MODELS FOR BEAM COHERENT OSCILLATION}
\end{center}

Two theoretical models, the rigid-beam model and the linearized Vlasov equation 
with the single-mode approximation, were examined with the HERA experiment. For 
the rigid-beam model, the coherent tunes were calculated with two different 
methods: assuming a Gaussian distribution as the equilibrium distribution of 
the $e^+$ beam [Eqs. (A10) and (A14) in Appendix A.1] or using the non-Gaussian 
quasi-equilibrium distribution of the $e^+$ beam obtained in the beam-beam 
simulation (Appendix A.2). In the case of the Gaussian distribution, the beam 
sizes used in Eq. (A14) are of the experimental measurement or the beam-beam 
simulation. The small differences in the beam sizes measured in the experiment
or calculated from the simulation (see Fig. 1) made little difference in 
Eq. (A14). For the linearized Vlasov equation, the horizontal coherent tunes
were obtained by solving the initial-value problem of the linearized Vlasov
equation with the single-mode approximation in the horizontal plane. The 
details of the calculations are in Appendix A and B.

Tables II and III list the ratio of the coherent and incoherent beam-beam 
tune shifts of the $e^+$ beam, $\Delta\nu_{e,x}/(2\xi_{e,x})$ and 
$\Delta \nu_{e,y}/(2\xi_{e,y})$, calculated in the beam simulation or 
calculated with the theoretical models for the cases of $\beta_{e,y}=1.0$ and 
4.0 m, where $\Delta\nu_{e,x}$ and $\Delta\nu_{e,y}$ are the horizontal and
vertical coherent beam-beam tune shift of the $e^+$ beam. The experimental 
measurement at $\beta_{e,y}=4.0$ m is also included in Table II. The 
significant discrepancy between the results of the models and the results of 
the experiment/simulation shows that the theoretical models are inconsistent 
with the experiment and simulation. Note that the rigid-beam model with the 
beam distribution from the simulation did a little better than that with 
Gaussian distribution. To further examine the failure of the theoretical 
models, a beam-beam simulation was conducted for the case of 
$\beta_{e,y} = 1.0$ m but with only one tenth of the $p$-bunch current used 
in the experiment, i.e. $\xi_{e,y}=0.0068$. The result is listed in Table IV 
and shows that the rigid-beam model with either the Gaussian distribution or 
the distribution obtained from the simulation are in a good agreement with 
the beam simulation. In fact, the quasi-equilibrium distribution of the $e^+$ 
beam in this case is very close to a Gaussian. This re-confirms many previous 
studies that the rigid-beam model is correct for unsymmetrical beam-beam 
interactions with relatively small beam-beam parameter. 

In the case of the linearized Vlasov equation, as shown in Tables II-IV, the
calculation yielded wrong results no matter how small the beam-beam parameter 
is. In the calculation with the Vlasov equation, several approximations 
were employed. Among them, the linearization, the one-dimensional beam, and 
the single-mode approximation are the three major approximations that cannot 
be directly justified by the experimental observations (see Appendix B). 
The linearization of the Vlasov equation should not play the leading role 
in its failure for the unsymmetrical beam-beam interaction since the 
linearization should be valid for a relatively small beam-beam parameter. To 
verify the validity of the one-dimensional approximation, we did a series 
of beam-beam simulations for the case of $\beta_{e,y}=1.0$ m, 
$\xi_{e,y}=0.068$ or 0.0068, but with a different beam aspect ratio
($\sigma_{y,e}/\sigma_{x,e}$) ranging from 0.063 
to 0.120. In all these cases, the horizontal coherent beam-beam tune shift 
was found to be very similar to that of the HERA experiment. Note that in 
the HERA experiment and in all the cases in Tables II-IV, 
$\sigma_{y,e}/\sigma_{x,e}=0.126$. This indicates that the horizontal 
coherent beam-beam tune shift is not sensitive to the aspect ratio of the 
beam. Similar phenomenon has also been observed in the symmetrical case of 
beam-beam interactions \cite{Yokoya}. The large discrepancy in the coherent
beam-beam tune shifts calculated with the linearized Vlasov equation as
compared to that obtained from the experiment/simulation is apparently not 
due to the approximation of one-dimensional beam. With the single-mode 
approximation, on the other hand, the only oscillation mode of the beam 
distribution that was kept in the calculation is the $m=1$ mode in Eq. (B6). 
As shown in Fig. 6, the equilibrium distribution of the $e^+$ beam 
[$f_{1,0}(I,\phi)$ in Eq. (B4)] depends strongly on $\phi$. In this case, 
the dominant mode of the equilibrium distribution is the $m=2$ mode which 
has a $\phi$-dependence of $e^{i2\phi}$. As the $m=2$ mode could be a 
dominant intrinsic mode of the system, the single-mode approximation could 
be inconsistent with the dynamics of the beam. In order to study the effect 
of mode couplings in the linearized Vlasov equation, we have derived the 
eigenvalue equation of the linearized Vlasov 
equation that is similar to Eq. (B13) but includes high-order modes. After 
truncating higher-order modes at $m=m_p$, ${\bf M}$ in Eq. (B13) becomes a
$(2m_p+1)(2l_p+2)\times(2m_p+1)(2l_p+2)$ matrix where $l_p$ is the number 
of grids on the mesh of the action space and ${\bf M}_1$ and ${\bf M}_2$ in 
Eq. (B14) are no longer diagonal matrices (see Appendix B). We have, however, 
failed in obtaining a set of orthogonal eigenvectors for the eigenvalue 
equation of the linearized Vlasov equation because ${\bf M}$ is a singular 
(ill-conditioned) matrix when the mode couplings are included. Similar
problem has also been encountered when including the mode coupling in the 
case of symmetrical beam-beam interactions. 

\begin{center}
{\bf V. COHERENT VERSUS INCOHERENT BEAM-BEAM TUNE SHIFT}
\end{center}

In the symmetrical case of beam-beam interactions, the ratio of the coherent and 
incoherent beam-beam tune shift is approximately a constant of 1.2 for round 
beams and 1.3 for flat beams \cite{Yokoya}. Note that with symmetrical beam-beam
interactions, the beam-size growths of the two beams are symmetrical and the 
beam distributions are usually close to a Gaussian when the beam-beam parameter 
is below the threshold of the onset of coherent beam-beam instability
\cite{Shi,Jin}. The coherent beam-beam tune shift depends linearly on the 
beam-beam parameter in a fairly large range of beam-beam parameter \cite{Yokoya}. 
On the contrary, in the HERA experiment the ratio of the coherent and incoherent
beam-beam tune shift of the $e^+$ beam decreases monotonically with the increase 
of the beam-beam parameter as shown in Fig. 7. This different characteristic of 
the coherent beam-beam tune shift stems mainly from the mismatch in the 
equilibrium distributions of two unsymmetrical colliding beams. In the HERA 
experiment, the beam size of the $e^+$ beam at the IPs is slightly larger than 
that of the $p$ beam initially without collision (see Table I). When the beam-beam 
parameter is small such as in the case of $\xi_{e,y}=0.0068$, the beam-size growth 
is insignificant and the beams are close to the Gaussians. The ratio of the 
coherent and the incoherent beam-beam tune shift of the $e^+$ beam is simply 
determined by the mismatch in the initial beam sizes without collision as 
described by Eq. (A14) of the rigid-beam model. When the beam-beam parameter is 
large, on the other hand, the beam-size growth of the $e^+$ beam is significant 
(see Fig. 1) and dominates the beam-size mismatch. Moreover, the particle
distribution of the $e^+$ beam deviates significantly from a Gaussian 
distribution (see Fig. 5). In this situation, the size of the $e^+$ beam is much 
larger than that of the $p$ beam and a large number of $e^+$ particles escape to 
the beam tails that is away from the core of the $p$ beam. The coherent beam-beam 
tune shift of the $e^+$ beam thus becomes much smaller than that in the case of 
two matched Gaussian beams. Note that for two matched Gaussian beams, the ratio 
of the coherent and incoherent beam-beam tune shift is 0.5 for unsymmetrical 
beam-beam interactions. As the beam-size growth and the resultant beam-size
mismatch increase with the beam-beam parameter, the ratio of the coherent and 
incoherent beam-beam tune shift of the $e^+$ beam decreases with the beam-beam 
parameter. In general, the equilibrium states of the Vlasov equation in Eq. (B4) 
are different for two unsymmetrical colliding beams due to the asymmetry of 
beam-beam perturbations. The mismatch in the distributions as well as the beam 
sizes is, in principle, independent of the initial states of the beam 
distributions and the initial beam-size mismatch as long as the considered 
equilibrium state of the Vlasov equation in Eq. (B4) is well isolated and is 
the only one that is close to the initial beam distributions. The beam-size 
mismatch in the unsymmetrical case of beam-beam interactions is therefore 
intrinsic and unavoidable especially when the beam-beam parameter is larger. 
Moreover, this beam-size mismatch increases with the strength of beam-beam 
perturbations. The functional dependence of the coherent beam-beam tune shift 
on the beam-beam parameter in Fig. 7 is therefore a general characteristic of 
unsymmetrical beam-beam interactions.

Figure 7 also plots $\Delta \nu_{e,y}/(2\xi_{e,y})$ as a function of $\xi_{e,y}$ 
calculated based on the rigid-beam model. It shows that when the beam-beam
parameter is large, the rigid-beam model results in a similar  
$\xi_{e,y}$-dependence of $\Delta \nu_{e,y}/(2\xi_{e,y})$ even though it 
overestimates the coherent beam-beam tune shift. Note that in the rigid-beam 
model, this $\xi_{e,y}$-dependence is due to the beam-size mismatch as given 
in Eq. (A14). At a large $\xi_{e,y}$, the beam-size mismatch is dominated by 
the beam-size growth of the $e^+$ beam. This suggests that the result of the 
rigid-beam model in Eq. (A14) provides an approximation of the functional 
dependence of the coherent beam-beam tune shift to the intrinsic beam-size 
mismatch in unsymmetrical beam-beam interactions with large beam-beam parameters. 

\begin{center}
{\bf VI. CONCLUSION}
\end{center}

The coherent beam-beam tune shift was studied in the case of unsymmetrical
beam-beam interactions where the two beams have very different beam-beam
parameters and betatron tunes. The results of a self-consistent beam-beam
simulation, the rigid-beam model, and the linearized Vlasov equation were
compared with the beam measurement in the HERA 2000 Beam Study. Remarkable
agreement was found between the beam simulation and the HERA experiment in 
a wide range of and, especially, at very large beam-beam parameters of the
lepton beam. The rigid-beam model was found to be only correct when the 
beam-beam parameter is small. The result of the linearized Vlasov equation 
with the single-mode approximation is inconsistent with the result of the 
beam experiment/simulation in either cases of large or small beam-beam
parameter. The failure of the linearized Vlasov equation could be due to the 
single-mode approximation used in solving the linearized Vlasov equation. 
A study of the dynamics of the beam distribution showed that the high-order
modes are important to the beam dynamics in this situation. An attempt to
include high-order modes in the calculation has, however, not been successful
because of the difficulty in finding a set of orthogonal eigenvectors for the
linearized Vlasov equation. Recently, efforts have been made to include the
angle-dependence of beam distributions in an expansion of the Vlasov equation
\cite{Schmekel}. More studies are needed for a relevant solution of the 
linearized Vlasov equation for the unsymmetrical beam-beam interaction. 
Currently, the numerical simulation is the only reliable approach for an 
prediction of the coherent beam-beam tune shift in this situation.

One interesting phenomenon observed in this study is the very small coherent 
beam-beam tune shift in this unsymmetrical case of beam-beam interactions. It 
was found that the ratio of the coherent and incoherent beam-beam tune shift 
of the weak lepton beam in HERA decreases from 0.3 to 0.02 as the incoherent 
beam-beam tune shift increases from 0.01 to 0.54. On the contrary, in the 
symmetrical case of beam-beam interactions, this ratio maintains approximately 
a constant of 1.2 for a round beam or 1.3 for a flat beam in a large range 
of beam-beam parameter. The reason for this different characteristic of the 
coherent beam-beam tune shift is the intrinsic beam-size mismatch between 
two unsymmetrical colliding beams due to the difference in the equilibrium 
distributions of the two beams. This intrinsic mismatch in the beam 
distributions due to beam-beam interactions becomes more pronounced as the 
strength of beam-beam perturbations increases. The ratio of the coherent and
incoherent beam-beam tune shift decreases, in general, with the increase of 
beam-beam parameters in the unsymmetrical case of beam-beam interactions.
   
\begin{center}
{\bf APPENDIX A. COHERENT BEAM-BEAM TUNE SHIFT FROM RIGID-BEAM MODEL}
\end{center}

The coherent beam-beam tune shift in the rigid-beam model has been studied 
before for the unsymmetrical beam-beam interaction. Previous studies are, 
however, limited to the special cases where either the two beams have the 
same \cite{Hofmann} or very different \cite{Georg99} beam-beam parameters.
Moreover, all of them assumed that the beams are Gaussian. In the following, 
a general formula of the rigid-beam model is derived that can be apply to 
any case of unsymmetrical beam-beam interaction with either Gaussian or 
non-Gaussian beams.

Let $\rho_i\left(\vec{r}, \theta\right)$ be the distribution of beam $i$ 
in the normalized configuration space where $i=1$ or 2, $\vec{r}=(x,y)$ 
are the normalized coordinates of the transverse space, and $\theta$ 
the azimuthal angle associated with the path length along the closed 
orbit. The beam centroid in the normalized space can be calculated by
$\vec{R}_i=\int \vec{r}\rho_i\left(\vec{r},\theta\right)d\vec{r}$. 
Considering a linear lattice with one IP, the transverse motion of the 
beam centroid can be described by
$$
\frac{d^2\vec{R}_i}{d\theta^2}+{\bf \Omega}_i\cdot\vec{R}_{i}
=(-1)^{i+1}\;\lambda_i \vec{F}\;\left[2\pi\sum_n\delta(\theta-2\pi n)\right] \;. \eqno{(A1)} 
$$
In Eq. (A1), ${\bf \Omega}_i$ is a $2\times 2$ diagonal matrix with 
$({\bf \Omega}_i)_{11}=\nu_{i,x}^2$ and $({\bf \Omega}_i)_{22}=\nu_{i,y}^2$,
where $(\nu_{i,x},\;\nu_{i,y})$ are the fractional parts of the betatron 
tunes of the lattice for beam $i$. The main part of the beam-beam kick in 
Eq. (A1) is
$$
\vec{F}= \int\limits_{-\infty}^{+\infty}\int\limits_{-\infty}^{+\infty}
\rho_1\left(\vec{r}_1,\theta\right)\rho_2\left(\vec{r}_2,\theta\right) 
\vec{G}\left(\beta_{1,x}^{1/2}x_1-\beta_{2,x}^{1/2}x_2 \;, \;
 \beta_{1,y}^{1/2}y_1-\beta_{2,y}^{1/2}y_2 \right)d\vec{r}_1d\vec{r}_2  
\eqno{(A2)}
$$
where $\vec{G}(x,y)=\vec{r}/r^2$ is the Green Function of the beam-beam
interaction and $(\beta_{i,x},\;\beta_{i,y})$ are the horizontal and
vertical beta-function of beam $i$ at the IP, respectively. The strength 
of the beam-beam kick for the horizontal component of Eq. (A1) is 
$$
\lambda_i=\lambda_{i,x}=\frac{a_iN_j}{\pi\gamma_i}\nu_{i,x}\beta^{1/2}_{i,x}
$$ 
where $i=$1 or 2, $j=$1 or 2, but $i\neq j$; $N_j$ is the number of particles 
per bunch of beam $j$; $a_i$ the classical radius of the particle in beam $i$;
and $\gamma_i$ the Lorentz factor of beam $i$. With the definition 
of the beam-beam parameters in Eq. (1), this kick strength can be written as
$$
\lambda_{i,x}=\frac{2\nu_{i,x}\;\xi_{i,x}}{\beta_{i,x}^{1/2}}\;
\sigma_{j,x}(\sigma_{j,x}+\sigma_{j,y}) \eqno{(A3)}
$$
where $(\xi_{i,x},\;\xi_{i,y})$ and $(\sigma_{i,x},\;\sigma_{i,y})$ are 
the beam-beam parameters and the rms beam sizes at the IP of beam $i$. For 
the vertical component of Eq. (A1), $\lambda_i=\lambda_{i,y}$ can be easily 
obtained by exchanging $x$ and $y$ in Eq. (A3). 

In the rigid-beam model of the coherent beam-beam oscillation, the shapes of 
the particle distributions in phase space are assumed to not change with time 
during the beam oscillation while the centers of the distributions oscillate 
with the beams' coherent tunes. The distribution during the beam oscillation 
is thus assumed to be $\rho_i(\vec{r},\theta)=\rho_{0i}(\vec{r}-\vec{R}_i)$,
where $\rho_{0i}(\vec{r})$ is the equilibrium distribution when 
the beam is centered at the closed orbit. In general, $\vec{F}$ in Eq. (A2) 
is a function of moments of phase-space variables and the time-dependence of 
$\vec{F}$ is implicitly through all the moments. (To have a better picture
of this, one may consider a moment expansion of beam particle distributions 
in phase space). With the rigid-beam approximation, $\vec{F}$ depends on 
the lowest-order moments $\vec{R}_1$ and $\vec{R}_2$ only,
$$
\vec{F}=\int\limits_{-\infty}^{+\infty}\int\limits_{-\infty}^{+\infty}
\rho_{01}\left(\vec{r}_1-\vec{R}_1\right)
\rho_{02}\left(\vec{r}_2-\vec{R}_2\right) 
\vec{G}\left(\beta_{1,x}^{1/2}x_1-\beta_{2,x}^{1/2}x_2 \;, \;
 \beta_{1,y}^{1/2}y_1-\beta_{2,y}^{1/2}y_2 \right)d\vec{r}_1d\vec{r}_2 
\eqno{(A4)}
$$
Note that the rigid-beam model may fail in cases where the variation of the 
distributions is important during the beam coherent oscillation. To find 
the oscillation frequencies of $\vec{R}_i$, one can average the beam-beam 
kick in Eq. (A1) over one turn ($2\pi$ in longitudinal direction) and expand
$\vec{F}$ into a Taylor series of $\vec{R}_i$. Keeping only the linear terms 
of $\vec{R}_i$, Eq. (A1) becomes a coupled four-dimensional harmonic 
oscillator, 
$$ 
\frac{d^2\vec{R}_i}{d\theta^2}+{\bf \Omega}_i\cdot\vec{R}_i=(-1)^{i+1}\lambda_i
\left({\bf A}_1\cdot \vec{R}_1 + {\bf A}_2 \cdot\vec{R}_2 \right)
 \eqno{(A5)}
$$
where $i=1$ or 2, and ${\bf A}_i$ are $2\times 2$ matrices with
$$
{\bf A}_i=\left.\frac{\partial \vec{F}}{\partial \vec{R}_i}
\right|_{\vec{R}_1=0,\vec{R}_2=0} \;. \eqno{(A6)}
$$
If both the beams are mirror symmetric with respect to the horizontal and 
vertical plane, ${\bf A}_i$ are diagonal matrices and the horizontal and 
vertical coherent oscillation are decoupled. The two eigenfrequencies for 
the coherent oscillation in the horizontal plane can then be solved as
$$
\nu_{\pm}=\frac{1}{\sqrt{2}} \sqrt{\omega^2_{1} + \omega^2_{2}
\pm \sqrt{\left(\omega^2_{1}-\omega^2_{2} \right)^2
+16\nu_{1,x}\nu_{2,x}\delta\omega_{1}\delta\omega_{2}}} \eqno{(A7)}
$$
where 
$$
\begin{array}{l}
\omega^2_{i}=\nu_{i,x}^{2}-2\nu_{i,x}\delta\omega_{i} \\
\delta \omega_{i}=\lambda_{i,x} \left({\bf A}_i\right)_{11} 
/\left(2\nu_{i,x}\right) 
\end{array} \eqno{(A8)}
$$
for $i=1$ or 2. If $\delta\omega_i << \nu_{i,x}$ and in Eq. (A7)
$$ 
(\omega^2_1-\omega^2_2 )^2 >>
16 \nu_{1,x}\nu_{2,x}\delta\omega_1\delta\omega_2  \eqno{(A9)}
$$
then the coherent frequencies of the two beam are
$$
\left\{
\begin{array}{l}
\nu_{+}=\omega_1=\nu_{1,x}-\delta\omega_1 \\
\nu_{-}=\omega_2=\nu_{2,x}-\delta\omega_2
\end{array} \right . \eqno{(A10)}
$$ 
where $\nu_{+}$ and $\nu_{-}$ are the horizontal coherent tunes of beam 1 
and beam 2, respectively. Note that in the unsymmetrical case of beam-beam 
interactions, the two eigenfrequencies do not correspond to the so called 
$0$- (or $\sigma$-) and $\pi$-modes of symmetrical beam-beam interactions. 
The condition in Eq. (A9) can be further simplified as
$$ 
|\nu_{1,x}-\nu_{2,x}|>>\sqrt{\delta\omega_1\delta\omega_2} \;.
\eqno{(A11)}
$$
Therefore, if the difference of the lattice tunes is much larger than the 
geometric average of the coherent beam-beam tune shifts of the two beams, the 
coherent beam-beam tune shifts can simply be calculated with Eq. (A10). Note
that in the HERA experiment this condition was fulfilled. For the case of 
strong-weak beam-beam interactions such as $\xi_{1,x}>>\xi_{2,x}$, one can 
expand $\nu_{\pm}$ in terms of $\lambda_{2,x}/\lambda_{1,x}$. Keeping only 
the dominant term in the coherent beam-beam tune shifts yields
$$
\left\{
\begin{array}{l}
\nu_{+} = \nu_{1,x}-\delta\omega_1   \\
\nu_{-} = \nu_{2,x}-
\frac{\nu^2_{1,x}-\nu^2_{2,x}}
{\nu^2_{1,x}-\nu^2_{2,x}-2\nu_{1,x}\delta\omega_1}\;\delta\omega_2
\end{array} \right . \eqno{(A12)}
$$
where $\nu_{+}$ and $\nu_{-}$ are the coherent frequencies of the weak (beam 
1) and strong (beam 2) beam, respectively. In the first equation of Eq. (A12), 
since the zeroth-order term ($\delta\omega_1$) of $\lambda_{2,x}/\lambda_{1,x}$ 
exists and dominates the coherent beam-beam tune shift of beam 1, the 
first- or higher-order terms were neglected. In the second equation of Eq. (A12),
on the other hand, the zeroth-order term is zero and the first-order term was 
thus kept. Note that if the lattice tunes of the two beams are very different, 
Eq. (A12) is equivalent to Eq. (A10). If the denominator in Eq. (A12), 
$\nu_{1,x}^2-\nu_{2,x}^2-2\nu_{1,x}\delta\omega_1$, is small, an analysis of 
the higher order terms shows that the expansion in terms of 
$\lambda_{2,x}/\lambda_{1,x}$ is no longer accurate and the coherent tunes 
have to be calculated by using Eq. (A7). The two coherent frequencies in 
the vertical plane can be easily obtained by exchanging $x$ and $y$ and 
changing $\left({\bf A}_i\right)_{11}$ with $\left({\bf A}_i\right)_{22}$ in 
Eqs. (A7)-(A12). As shown in our study (see Section IV), this approach of 
the rigid-beam model is quite good in the case of unsymmetrical beam-beam 
interactions with a small beam-beam parameter.

\noindent
{\bf A.1. Gaussian Beams}

In the case that $\rho_{0i}$ are Gaussian distributions, matrix ${\bf A}_i$ 
in Eq. (A6) can be calculated analytically with Eq. (A4) as
$$
\left({\bf A}_i\right)_{11}=\frac{ \beta_{i,x}^{1/2}}
{\Sigma_x(\Sigma_x+\Sigma_y)} \eqno{(A13)}
$$
where $\Sigma_x=\sqrt{\sigma_{1,x}^2+\sigma_{2,x}^2}$ and  
$\Sigma_y=\sqrt{\sigma_{1,y}^2+\sigma_{2,y}^2}$. The matrix element of
$\left({\bf A}_i\right)_{22}$ can be simply obtained by exchanging $x$ and 
$y$ in Eq. (A13). Substituting Eqs. (A2) and (A13) into Eq. (A8) yields
$$
\delta\omega_i = \xi_{i,x}\frac{(\sigma_{j,x}+\sigma_{j,y})\sigma_{j,x}}
{(\Sigma_x+\Sigma_y)\Sigma_x} \eqno{(A14)}
$$
where $i=1$ or 2; $j=1$ or 2; but $j\neq i$. When $i=1$, Eq. (A14) gives 
the coherent beam-beam tune shifts of the weak beam [beam 1 in Eq. (A12)] 
obtained previously by Hoffstaetter for the strong-weak case of beam-beam 
interactions \cite{Georg99}. If $(\nu_{1,x},\nu_{1,y})=(\nu_{2,x},\nu_{2,y})$ 
and $(\sigma_{1,x},\sigma_{1,y})=(\sigma_{2,x},\sigma_{2,y})$, 
Eq. (A7) is reduced to the formula obtained by Hirata \cite{Hirata}. 
If $(\xi_{1,x},\xi_{1,y})=(\xi_{2,x},\xi_{2,y})$ and 
$(\sigma_{1,x},\sigma_{1,y})=(\sigma_{2,x},\sigma_{2,y})$, Eq. (A7) is 
reduced to the formula obtained by Hofmann \cite{Hofmann}. 

\noindent
{\bf A.2. Non-Gaussian Beams}

For non-Gaussian beams, especially the distributions obtained from beam-beam 
simulations such as that in Figs. 5 and 6, matrices ${\bf A}_i$ in Eq. (A6)
cannot be obtained analytically but can be calculated numerically by using 
Eq. (A4). The coherent frequencies can then be calculated with Eq. (A7) or 
directly from Eq. (A5) if the horizontal and vertical motion are coupled.

\begin{center}
{\bf APPENDIX B. COHERENT BEAM-BEAM TUNE SHIFT FROM LINEARIZED VLASOV EQUATION}
\end{center}

The use of the linearized Vlasov equation has been very successful for the 
coherent beam-beam tune shift in the case that two beams have the same or 
very close lattice tunes \cite{Yokoya}. In order to find the coherent 
beam-beam tune shift, one needs to identify the coherent frequencies from the 
eigenfrequencies of the linearized Vlasov equation. The linearized Vlasov 
equation, in principle, has infinite numbers of eigenfrequencies associated 
with infinite numbers of oscillation modes. For real beams, the number of the 
eigenfrequencies of the beam oscillation in transverse space is twice of the 
number of particles in a bunch. When two beams have the same lattice tunes, 
the coherent frequencies can be easily identified since the eigenfrequencies 
that correspond to the coherent frequencies are separated from the rest of 
the eigenfrequencies that form a continuous band (many close eigenfrequencies
lines) \cite{Yokoya}. Although, it is not very clear mathematically why this 
separation occurs. The situation becomes more complicated when two beams have 
very different lattice tunes. In this case, all the eigenfrequencies are in 
one or two continuous bands and the coherent frequencies cannot be identified 
by only solving the eigenfrequencies. In order to find the coherent 
frequencies in the HERA beam experiment, we will instead solve the 
initial-value problem of the linearized Vlasov equation for the coherent 
beam oscillation.

Consider only the horizontal motion (very flat beam) in a linear lattice with 
one IP. In terms of action-angle variable, the Hamiltonian for the betatron
motion of beam $i$ ($i=1$ or $2$) can be written as
$$ 
H_i(I,\phi,\theta) = H_{i,0}(I)+ U_i(I,\phi,\theta)
\left[2\pi\sum_n\delta(\theta-2\pi n)\right] \eqno {(B1)} 
$$ 
where $H_{i,0} = \nu_{i,x}I $ is the Hamiltonian associated with the betatron 
motion in the linear lattice and $U_i$ is the potential energy for the 
beam-beam interaction that can be written, for one-dimensional beams, as
$$ U_i(I,\phi,\theta)= U_i[f_j]=-2\frac{\xi_{i,x}\sigma_{j,x}^2}{\beta_{i,x}}
\int\limits_{0}^{2\pi}\int\limits_{0}^{\infty}f_j(I',\phi',\theta)
\ln\left(\sqrt{2\beta_{i,x}I}\sin\phi-\sqrt{2\beta_{j,x}I'}\sin\phi'\right)
dI'd\phi' \eqno{(B2)}
$$
where $i=1$ or 2, and $j=1$ or 2, but $i\neq j$. The action-angle variables 
are related to the normalized variables by $x=\sqrt{2I}\sin\phi$ and 
$p=\sqrt{2I}\cos\phi$. $f_i(I,\phi,\theta)$ is the particle distribution of 
beam $i$ in phase space and satisfies the Vlasov equation. For convenience,
we also define a functional $U_i[f_j]$ in Eq. (B2) for the potential integral. 
In Eq. (B2), $\ln(x-x')$ is the Green function for the potential of beam-beam 
interaction in one-dimensional space. If only the coherent 
beam-beam tune shifts are interested, one can get rid of the periodic 
$\delta$-function in the Hamiltonian in Eq. (B1) by average the beam-beam 
force over one turn. The Vlasov equation for $f_i$ can then be written as
$$
\frac{\partial f_i}{\partial\theta}+\nu_{i,x}\frac{\partial f_i}{\partial\phi}
=\{U_i,f_i\} 
\eqno{(B3)}
$$
where $\{ \;\}$ is the Poisson bracket. Assume that the beams have reached
equilibrium distributions $f_{i,0}$ that satisfy
$$
\nu_{i,x}\frac{\partial f_{i,0}}{\partial\phi} = \{U_{i,0},f_{i,0}\}  
\eqno{(B4)} 
$$
where $U_{i,0}(I,\phi)=U_i[f_{j,0}]$. Consider that beam $i$ experiences 
a small perturbation from its equilibrium distribution 
$\psi_i(I,\phi,\theta)=f_i(I,\phi,\theta)-f_{i,0}(I,\phi)$. The linearized 
equation for $\psi_i(I,\phi,\theta)$ can be obtained by subtracting Eq.(B4) 
from Eq. (B3) and neglecting the term $\left\{U_i[\psi_j],\;\psi_i\right\}$
which is higher-order in $\psi_i$ as
$$
\frac{\partial\psi_i}{\partial\theta} +  \nu_{i,x} \frac{\partial\psi_i} 
{\partial\phi} = \{U_{i,0},\psi_i\} + \{V_i,f_{i,0}\}. \eqno{(B5)}
$$  
where $V_i(I,\phi,\theta)=U_i[\psi_j]$. 

To solve Eq. (B5), one can convert it into a system of infinite numbers of 
coupled ordinary differential equations of modes by using Fourier 
transformation
$$
\psi_i(I,\phi,\theta) = \int\limits_{-\infty}^{\infty}d\nu
\sum\limits_{m=-\infty}^{\infty}\psi_{i,m}(I,\nu) e^{i(m\phi-\nu\theta)}
\eqno{(B6)}
$$
where $\nu$ is the oscillation frequency of the beams and $m$ numbers modes. 
The $m=1$ mode corresponds to the coherent dipole oscillation. To further
simplify the problem, one may use the single-mode approximation in which only 
the mode with $m=1$ is kept in the linearized Vlasov equation \cite{Yokoya}. 
It turns out that the use of single-mode approximation is not only a convenience 
but also a necessity. Without the single-mode approximation, no effective method 
is available for the general solution of Eq. (B5), except for simplified models 
as in Ref. \cite{Schmekel}. Substituting Eq. (B6) into Eq. (B5), multiplying 
$e^{-i\phi}$ and integrating $\phi$ over $2\pi$ on the both sides of Eq. (B5), 
and only keeping the $m=1$ mode yields
$$
\nu \bar{\psi}_i(I,\nu) = \nu_{i,x}\bar{\psi}_{i}(I,\nu)
+Q_i(I)\bar\psi_i(I,\nu)+\int\limits_{0}^{\infty}
G_i(I,I')\bar\psi_j(I',\nu)dI'
\eqno{(B7)}
$$ 
where $\bar{\psi}_i(I,\nu)=\psi_{i,1}(I,\nu)$, 
$$ Q_i(I)=\frac{1}{2\pi}\int\limits_{0}^{2\pi}\frac {\partial U_{i,0}}
{\partial I} d\phi \eqno{(B8)}
$$
and
$$ G_i(I,I')=\frac{\xi_{i,x}\sigma^2_{j,x}}{\pi\sqrt{2\beta_{i,x}I}}
\int\limits_0^{2\pi}\frac {e^{-i\phi}\sin\phi'}
{\sqrt{\beta_{i,x}I} \sin\phi-\sqrt{\beta_{j,x}I'}\sin\phi'}
\left(2I\cos\phi\frac{\partial f_{i,0}}{\partial I}
-\sin\phi\frac{\partial f_{i,0}}{\partial \phi} 
\right)d\phi'd\phi
\eqno{(B9)}
$$
If the equilibrium distributions are independent of $\phi$ such as for 
Gaussian beams, the imaginary term of $G_i(I,I')$ is zero. 
Otherwise, this imaginary term contributes a damping to the linearized 
Vlasov equation when $\bar{\psi}_i$ is stable or an excitation when 
$\bar{\psi}_i$ is unstable. If the equilibrium distributions $f_{i,0}$ 
are Gaussian, with a similar algebraic treatment in Ref. \cite{Yokoya}, 
the integrals in Eqs. (B8) and (B9) can be calculated analytically as 
$$
Q_i(I)= -\frac{\xi_{i,x}\sigma^2_{j,x}}{\beta_{i,x}I} 
\left(1-e^{-\beta_{i,x}I/\sigma^2_{j,x}}\right)
\eqno{(B10)}
$$
$$
G_i\left(I,I'\right)=\xi_{i,x} \; r_{ij} \; e^{-\left(z_i+z_j'\right)/2}
\left[\frac{\min\left(z_i,\; r_{ij}z_j' \right)}
{\max\left(z_i,\; r_{ij}z_j' \right)} \right]^{\frac{1}{2}} \eqno{(B11)}
$$
where $i=1$ or 2; $j=1$ or 2; but $i\neq j$. $z_i=\beta_{i,x}I/\sigma_{i,x}^2$,
$z'_j=\beta_{j,x}I'/\sigma_{j,x}^2$, and $r_{ij}=\sigma_{j,x}^2/\sigma_{i,x}^2$.

\medskip\noindent
{\bf B.1. Eigenfrequencies and Eigenvectors of Linearized Vlasov Equation}
\medskip

To further proceed with Eq. (B7), one may discretize the action space ($I$) 
into a mesh and solve the equation on the grids \cite{Meller,Yokoya}. Let 
$I= l\Delta I$ where $\Delta I$ is the grid size; $l=0$, 1, 2, ... $l_p$;
and $l_p\Delta I$ is the size of the mesh. Since the distributions decay to 
zero quickly as $I$ increases, a mesh that covers several $\sigma_{i,x}$ is 
good enough for a calculation of the coherent frequency. In order to have an 
accurate frequency for the lattice tune in the eigenfrequencies of Eq. (B7),
however, the mesh has to be large enough so that the incoherent beam-beam 
tune shift at $I=l_p\Delta I$ is negligible. In this study, we therefore used 
$l_p\Delta I=160 \;\epsilon_{i,x}$ and $\Delta I = 0.05\;\epsilon_{i,x}$, 
where $\epsilon_{i,x}$ is the normalized emittance of beam $i$. 
Let $\bar{\psi}_{i}(l\Delta I,\nu)=\bar{\psi}_{il}(\nu)$. Eq. (B7) can 
then be converted into a system of linear algebraic equations on the mesh,
$$
\nu \bar{\psi}_{il} = \nu_{i,x}\bar{\psi}_{il}
+ Q_i\left(l\Delta I\right)\bar{\psi}_{il}+ \Delta I \sum\limits_{k=0}^{l_p}
  G_i\left(l\Delta I,k\Delta I\right)\bar{\psi}_{jk} \; .
\eqno{(B12)}
$$
that leads to an eigenvalue problem
$${\bf M}\vec{V} = \nu \vec{V} \eqno{(B13)}$$ 
where 
$$
\vec{V} = \left(\bar{\psi}_{10}, \;\bar{\psi}_{11},\;\cdot\cdot\cdot,
 \;\bar{\psi}_{1l_p},\;\bar{\psi}_{20}, \;\bar{\psi}_{21},\;\cdot\cdot\cdot,
 \;\bar{\psi}_{2l_p} \right)^T  \;.
$$
${\bf M}$ is a $2(l_p+1)\times 2(l_p+1)$ matrix
$$
{\bf M} = \left (
\begin{array}{ll} {\bf M}_1 & {\bf O}_1 \\ {\bf O}_2 & {\bf M}_2 \\ 
\end{array} \right ) \eqno {(B14)}
$$
where ${\bf M}_1$, ${\bf M}_2$, ${\bf O}_1$, and ${\bf O}_2$ are 
$(l_p+1)\times (l_p+1)$ matrices. Because of the single-mode approximation,
${\bf M}_1$ and ${\bf M}_2$ are diagonal matrices with the diagonal elements
$$
\left({\bf M}_i\right)_{kk}=\nu_{i,x}-Q_i\left((k-1)\Delta I\right) 
\eqno{(B15)}
$$ 
and the elements of ${\bf O}_i$ are
$$
\left({\bf O}_i\right)_{kl}=G_i\left((k-1)\Delta I,\;(l-1)\Delta I\right)
\eqno{(B16)}
$$
where $k=1, ..., (l_p+1)$; $l=1, ..., (l_p+1)$; and $i=1$ or 2. If the 
equilibrium distributions are Gaussian, all these matrix elements in 
Eqs. (B15) and (B16) can be calculated analytically by using Eqs. (B10) and 
(B11). In the case of the HERA experiment, the equilibrium distribution of 
the $p$ beam is still very close to a Gaussian but the $e^+$ beam is no 
longer a Gaussian beam (see Figs. 5 and 6). Let beam 1 and 2 be the $e^+$ 
and $p$ beam, respectively. ${\bf M}_1$ and ${\bf O}_2$ can then be obtained 
analytically. The matrix elements of ${\bf M}_2$ and ${\bf O}_1$, on the 
other hand, have to be calculated numerically by using Eqs. (B8) and (B9) 
with the quasi-equilibrium distribution of the $e^+$ beam obtained from the 
beam simulation.

With the eigenvalue equation in Eq. (B13), the eigenfrequencies and a set of 
orthogonal eigenvectors for the linearized Vlasov equation can be found
numerically. If the two beams have the same lattice tune, the eigenfrequencies
of Eq. (B13) are identical to that obtained in Ref. \cite{Yokoya}. Fig. 8a 
is an example of eigenfrequencies of Eq. (B13) for the case of 
$\nu_{1,x}=\nu_{2,x}$ and $\xi_{1,x}=\xi_{2,x}$. It shows that in the
symmetrical case of beam-beam interactions the coherent frequency (the first 
frequency line from the left of Fig. 8a) is separated from the rest of the 
eigenfrequencies that form a continuous band. The width of the band equals 
the incoherent beam-beam tune shift. The coherent frequency can therefore be
easily identified in this case. Note that the coherent beam-beam tune shift 
calculated from this coherent frequency is the same as that in Ref. 
\cite{Yokoya}. The situation is more complicated when the two beams have very 
different lattice tunes. Fig. 8b plots the eigenfrequencies for the case of 
the HERA experiment. In this case, the eigenfrequencies are divided into two
groups, one for each beam. For the $e^+$ beam, the eigenfrequencies form a 
continuous band that starts at the lattice tune of the $e^+$ beam and has a 
width of the incoherent beam-beam tune shift of the $e^+$ beam. Because there 
were two IPs in the HERA experiment, the incoherent beam-beam tune shift in 
Fig. 8b is $2\xi_{e,x}=0.082$ for the $e^+$ beam. The characteristics of the
eigenfrequencies for the $p$ beam, in principle, is similar to that of the 
$e^+$ beam. Since the beam-beam parameter of the proton beam is very small
($\xi_{2,x}=\xi_{p,x}\sim 10^{-4}$), all the eigenfrequencies for the $p$ 
beam degenerate into a single line (the first line from the left of Fig. 8b) 
that corresponds to the lattice tune of the $p$ beam. In the case of very
unsymmetrical beam-beam interactions, therefore, the coherent frequencies
cannot be simply identified from the eigenfrequencies of the linearized 
Vlasov equation.

It should be noted that because of a very small beam-beam parameter of the 
$p$ beam, during the HERA experiment no coherent beam-beam tune shift was 
observed on the proton beam. The matrix elements of ${\bf O}_2$ are very 
small as compared with the diagonal elements of ${\bf M}_1$ and ${\bf M}_2$. 
${\bf O}_2$ can therefore be approximated as a zero matrix and the 
eigenfrequencies for the $e^+$ beam can be easily obtained from 
${\bf M}_1\vec{V}_1 = \nu \vec{V}_1 $, where $\vec{V}_1=\left(\bar{\psi}_{10}, 
\;\bar{\psi}_{11},\;\cdot\cdot\cdot,\;\bar{\psi}_{1l_p}\right)^{T}$ is the 
sub vector space associated with the $e^+$ beam. Since ${\bf M}_1$ is 
diagonalized, solving the eigenfrequencies and a set of orthogonal 
eigenvectors of ${\bf M}_1$ is trivial. The eigenfrequencies and 
eigenvectors obtained from ${\bf M}_1$ were found to be the same as that 
of Eq. (B13) in the sub vector space associated with the $e^+$ beam in this 
case.

\medskip
\noindent
{\bf B.2. Initial-Value Problem for Coherent Frequencies}
\medskip

Let $\left(\nu_1,\; \;\cdot\cdot\cdot, \; \nu_{l_p+1},\; 
\nu_{l_p+2},\; \;\cdot\cdot\cdot, \; \nu_{2l_p+2} \right)$ and
$\left(\vec{V}^{(1)},\;\cdot\cdot\cdot,\;\vec{V}^{(l_p+1)},\; \vec{V}^{(l_p+2)},
\;\cdot\cdot\cdot,\;\vec{V}^{(2l_p+2)}\right)$ be the eigenfrequencies 
and eigenvectors of the discretized and linearized Vlasov equation where 
${\bf M}\vec{V}^{(n)} = \nu_n\vec{V}^{(n)}$. In the HERA experiment, the
lattice tunes of the two beams are very different and, therefore, the two 
eigenfrequency bands of Eq. (B13) are well separated (see Fig. 8b). In this 
case, $\{\nu_n\}$ are the eigenfrequencies for the $e^+$ beam when 
$n=1, ..., l_p+1$ and the eigenfrequencies for the $p$ beam when 
$n=l_p+2, ..., 2l_p+2$. In the discretized action space, the perturbation of 
the beam distribution $\psi_1(I,\phi,\theta)$ and $\psi_2(I,\phi,\theta)$ 
can be represented as a vector, 
$$
\vec{\psi}(\theta)=\left(\;\psi_1(0,\phi,\theta),\;
\psi_1(\Delta I,\phi,\theta),\;\cdot\cdot\cdot,\; 
\psi_1(l_p\Delta I,\phi,\theta), \;\psi_2(0,\phi,\theta),\;\cdot\cdot\cdot,\; 
\psi_2(l_p\Delta I,\phi,\theta) \;\right)^T \;.
$$
With the single-mode approximation, the general solution of 
$\psi_1(I,\phi,\theta)$ and $\psi_2(I,\phi,\theta)$ can then be obtained from 
a superposition of the eigenvectors of the linearized Vlasov equation,
$$
\vec{\psi}(\theta) =\sum\limits_{k=1}^{2l_p+2} C_k\vec{V}^{(k)}
e^{i(\phi-\nu_k\theta)} \eqno{(B17)}
$$
where $\{C_k\}$ are constants and can be determined with an initial condition,
$\psi_1(I,\phi,0)$ and $\psi_2(I,\phi,0)$. Since $|C_k|^2$ is the oscillation
amplitude of the beam distributions with the frequency of $\nu_k$, the diagram 
of $|C_k|^2$ v.s. $\nu_k$ corresponds to the frequency spectrum of the coherent 
oscillation. The two peaks in the $|C_k|^2$-$\nu_k$ diagram, therefore, 
provides the coherent tunes when $\psi_1(I,\phi,0) \rightarrow 0$ and 
$\psi_2(I,\phi,0) \rightarrow 0$.

Consider a small kick that kicks beam 1 away from its equilibrium distribution 
$f_{1,0}(I,\phi)$, where $f_{1,0}(I,\phi)$ is known numerically from the 
beam-beam simulation. The initial perturbation of the beam distribution is 
$$
\psi_{1}(I,\phi,0)=f_{1,0}(x+x_0,p_x)-f_{1,0}(x,p_x)= 
\sum\limits_{m} g_m(I) e^{im\phi} \eqno{(B18)}
$$
and $\psi_2(I,\phi,0)=0$ where $x_0$ is the initial kick. With the single-mode 
approximation, $\psi_{1}(I,\phi,0) \simeq g_1(I) e^{i\phi}$ and 
$\vec{\psi}(0)=\vec{g}\;e^{i\phi}$ where
$\vec{g}=\left(\;g_1(0),\;g_1(\Delta I)\;,...,\;g_1(l_p\Delta I),\;
0\;,...,0\;\right)^T$. Note that the second half of the vector are all zero
because beam 2 is not kicked. On the other hand, from Eq. (B17)
$$
\vec{\psi}(0)=\left(\sum\limits_{k=1}^{2l_p+2} C_k \vec{V}_k\right)
e^{i\phi} =\left({\bf V}\vec{C}\right)e^{i\phi} \eqno{(B19)}
$$
where ${\bf V}$ is a $(2l_p+2)\times(2l_p+2)$ matrix of which the $i$th column 
is $\vec{V_i}$ and $\vec{C}=\left(C_1,C_2,...,C_{2l_p+2}\right)^{T}$. 
The coefficients $\{C_k\}$ can then be calculated from 
$\vec{C}= {\bf V}^{-1}\vec{g}$. It should be noted that the initial kick on 
the beam distributions in Eq. (B18) can be in any direction in phase space 
since the coherent frequency is the frequency of an infinitesimal oscillation. 
For near-integrable systems considered in this study, the phase-space 
region in the vicinity of the origin is integrable and only consists of 
invariant circles (tori). It is therefore isotropic. The coherent frequencies 
calculated were indeed found to be independent of the direction of the initial
kick.

Figure 8 plots the calculated $|C_k|^2$-$\nu_k$ diagrams for the symmetrical 
case of beam-beam interactions where $\nu_{1,x}=\nu_{2,x}$ and 
$\xi_{1,x}=\xi_{2,x}$ (Fig. 9a) and for the HERA experiment (Fig. 9b). In 
Fig. 9a, the peak with an arrow is the calculated coherent frequency that 
is the same as that in Fig. 8a. In Fig. 9b, the main peak indicates the 
calculated coherent frequency of the $e^+$ beam in the HERA experiment. The 
small peak in the lower right corner is the coherent frequency of the $p$ 
beam. Figure 10 plots the calculated coherent beam-beam tune shift of the 
$e^+$ beam as a function of initial kick $x_0$ for the case of the HERA 
experiment. It shows that the calculated coherent beam-beam tune shift
increases with the decrease of $x_0$ and converges as $x_0$ approaches
zero. This amplitude dependence of the coherent frequency is consistent 
with the beam simulation. Since the coherent frequency is the frequency of 
an infinitesimal oscillation, the convergence of the calculated coherent 
frequency at $x_0\rightarrow 0$ provides the wanted coherent frequency. 
As shown in our study (see Section IV), the linearized Vlasov equation 
with single-mode approximation is a valid approach for the coherent beam 
oscillation with symmetrical beam-beam interactions but not with 
unsymmetrical beam-beam interactions. 

\medskip
\noindent
{\bf ACKNOWLEDGMENTS}
\medskip

This work is supported by the US Department of Energy under Grant No. 
DE-FG02-04ER41288. The authors would like to thank the Center for Advanced 
Scientific Computing at the University of Kansas for the use of the 
Supercomputer. The authors thank F. Willeke for many stimulating
discussions. J. Shi would also like to thank DESY for the financial 
support for a trip to DESY at an early stage of our collaboration.

\newpage
\noindent
Table I. Some beam parameters used in HERA 2000 beam experiment. The 
horizontal size of the $e^+$ beam at the IPs is $\sigma_{e,x}=283$ $\mu$m. 
The horizontal and vertical size of the $p$ beam at the IPs are
$\sigma_{p,x}$=164 $\mu$m and $\sigma_{e,y}$=39.9 $\mu$m, respectively.

\vspace{0.1in}
\begin{center}
\begin{tabular}{|c|c|c|c|c|} \hline \hline
$\beta_{e,y}$ (m)  & I$_{e^+}$ (mA) & $\sigma_{e,y} (\mu m)$ 
& \hspace{0.15in} $\xi_{e,x}/\xi_{e,y}$ 
\hspace{0.15in} & $\xi_{p,x}/\xi_{p,y}$ $(10^{-4})$ \\ \hline 
1.0 & 19   & 35.8 & 0.041/0.068  & $2.54/1.40$ \\   
1.5 & 18   & 43.8 & 0.041/0.102  & $2.35/1.06$ \\
2.0 & 17   & 50.6 & 0.041/0.136  & $2.18/0.85$ \\  
3.0 & 3.5  & 62.0 & 0.041/0.204  & $0.43/0.14$ \\ 
4.0 & 2.6  & 72.0 & 0.041/0.272  & $0.31/0.09$ \\ \hline
\end{tabular}\end{center}

\vspace{0.5in}
\noindent
Table II. The coherent beam-beam tune shifts of the $e^+$ beam at 
$\beta_{e,y}=4.0$ m ($\xi_{e,y}=0.272$). ``Experiment'' and ``Simulation''
are the coherent tunes measured in the experiment and calculated in the beam 
simulation, respectively. ``Rigid-Real'' and ``Rigid-Gaussian'' are the
coherent tunes calculated by using the rigid-beam model with a Gaussian 
distribution and with the distribution from the simulation, respectively. 
``Vlasov Eq.'' is the coherent tunes calculated with the linearized Vlasov 
equation. 

\vspace{0.1in}
\begin{center}
        \begin{tabular}{|l|c|c|c|c|} \hline \hline
 $\beta_{e,y}=4.0$ m & $\nu_{e,x}$ &$\Delta \nu_{e,x}/(2\xi_{e,x})$
 &  $\nu_{e,y}$ & $\Delta \nu_{e,y}/(2\xi_{e,y})$ \\ \hline
 Experiment  & 0.1600 & 0.110 & 0.2330 & 0.024 \\ 
 Simulation  & 0.1605 & 0.104 & 0.2331 & 0.024 \\
 Rigid-Real  & 0.1555 & 0.164 & 0.2194 & 0.049 \\  
 Rigid-Gauss & 0.1531 & 0.194 & 0.2040 & 0.077 \\ 
 Vlasov Eq.  & 0.1123 & 0.69 & &  \\ \hline
        \end{tabular}
\end{center}

\newpage
\noindent
Table III. The same as Table II, but for $\beta_{e,y}=1.0$ m 
($\xi_{e,y}=0.068$).

\vspace{0.1in}
\begin{center}
        \begin{tabular}{|l|c|c|c|c|} \hline\hline
 $\beta_{e,y}=1.0$ m & $\nu_{e,x}$ &$\Delta \nu_{e,x}/(2\xi_{e,x})$
 &  $\nu_{e,y}$ & $\Delta \nu_{e,y}/(2\xi_{e,y})$ \\ \hline
 Simulation  & 0.1600 & 0.110 & 0.2172 & 0.212 \\ 
 Rigid-Real  & 0.1517 & 0.212 & 0.2074 & 0.284 \\  
 Rigid-Gauss & 0.1475 & 0.263 & 0.1996 & 0.341 \\ 
 Vlasov Eq.  & 0.121  & 0.58  & &             \\ \hline
        \end{tabular}
\end{center}

\vspace{0.5in}
\noindent
Table IV. The same as Table III, but with only one tenth of the $p$-bunch
current used in the experiment ($\xi_{e,y}=0.0068$).

\begin{center}
        \begin{tabular}{|l|c|c|c|c|} \hline\hline
 $\beta_{e,y}=1.0$ m & $\nu_{e,x}$ &$\Delta \nu_{e,x}/(2\xi_{e,x})$
 &  $\nu_{e,y}$ & $\Delta \nu_{e,y}/(2\xi_{e,y})$ \\ \hline
 Simulation  & 0.1672 & 0.224 & 0.2414 & 0.337 \\ 
 Rigid-Real  & 0.1668 & 0.263 & 0.2406 & 0.396\\ 
 Rigid-Gauss & 0.1668 & 0.268 & 0.2406 & 0.399 \\
 Vlasov Eq.  & 0.165  & 0.53  & &  \\ \hline
\end{tabular}
\end{center}

\newpage
\begin{center}
{\bf FIGURE CAPTIONS }
\end{center}

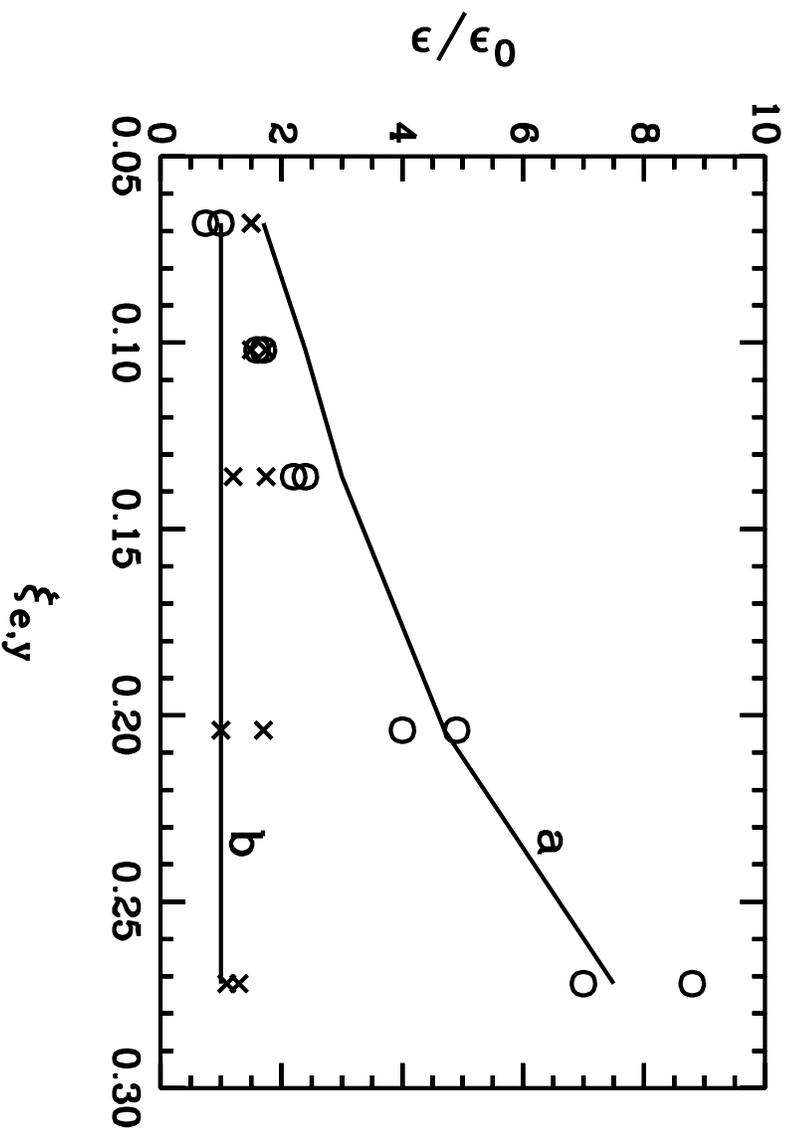
Figure 1. Emittance of the $e^+$ beam as a function of $\xi_{e,y}$. 
$\epsilon_0$ is the emittance without collision. Discrete points are from 
the experiment and continuous curves from the beam simulation. Circles and 
curve a is the vertical emittance. Crosses and curve b are the horizontal 
emittance. The two experimental data points at each $\xi_{e,y}$ where the 
measurement was performed correspond to the measurements at H1 and ZEUS.

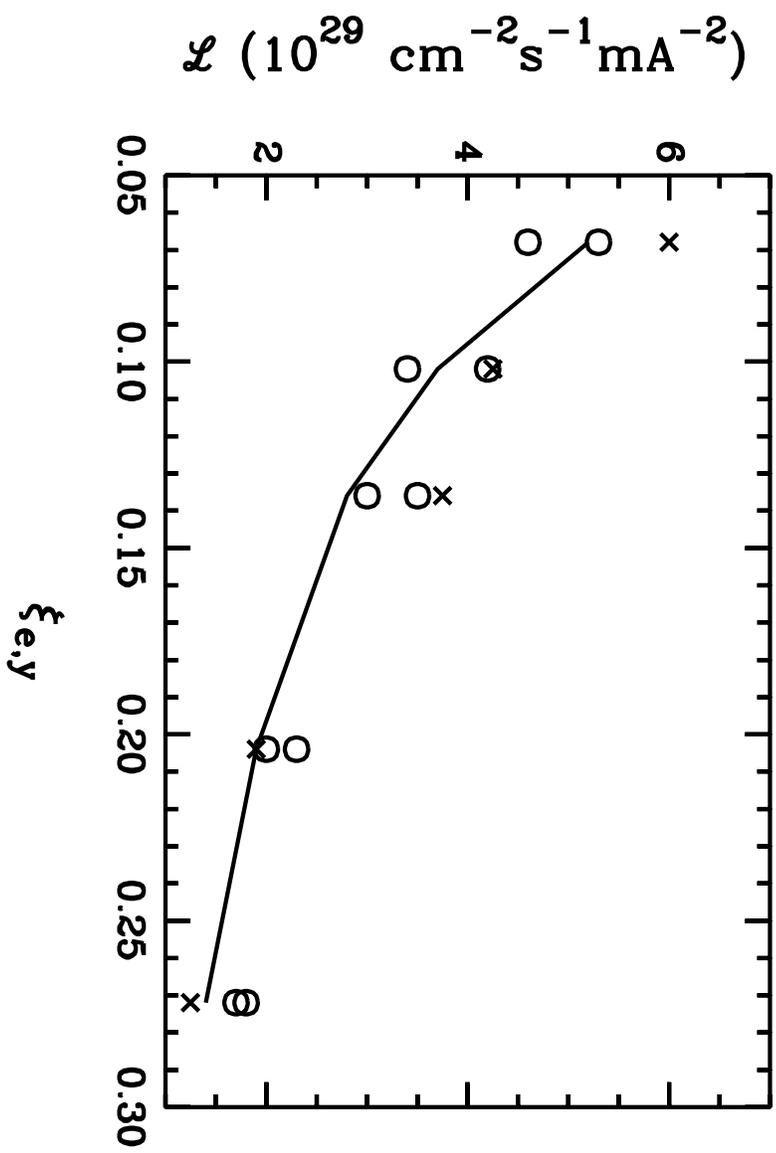
Figure 2. The specific luminosity as a function of $\xi_{e,y}$. Circles are 
from the experiment and continuous curves from the beam simulation. The two 
experimental data points at each $\xi_{e,y}$ where the measurement was 
performed correspond to the measurements at H1 and ZEUS. Crosses are the 
luminosity calculated with the measured emittance in Fig. 1 assuming
Gaussian beam distributions.

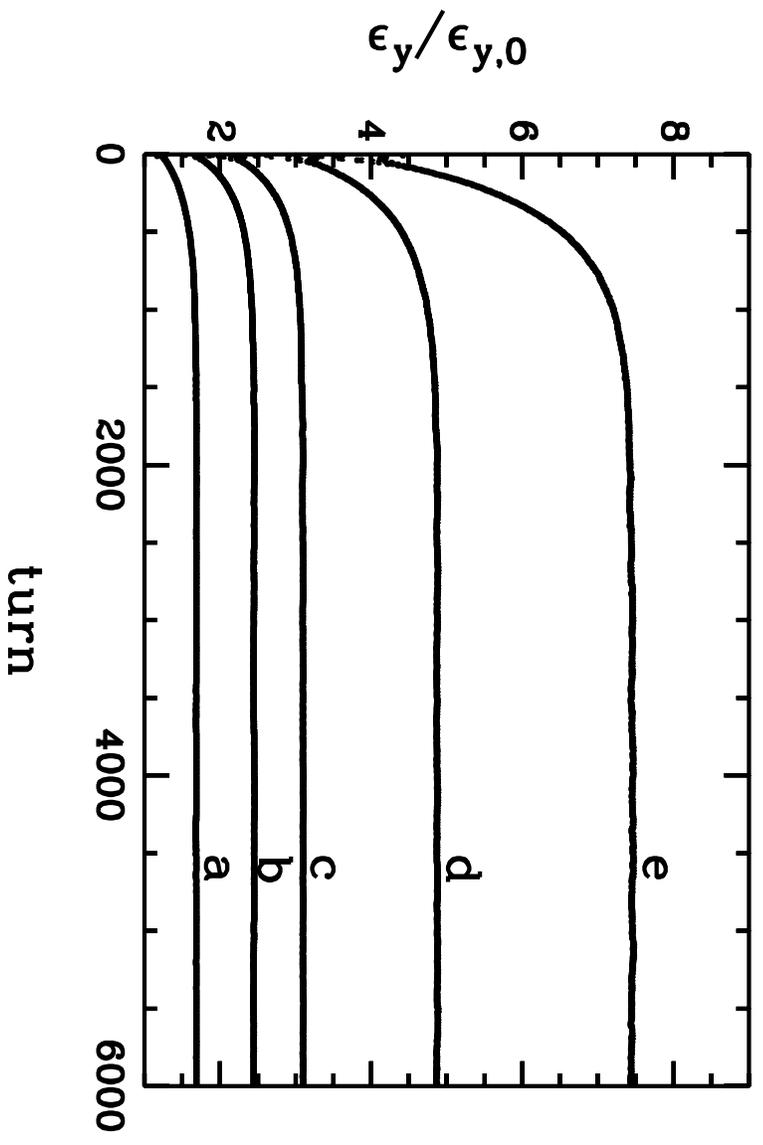
Figure 3. Evolution of the vertical emittance of the $e^+$ beam calculated 
with the beam simulation for the cases of HERA beam experiment at 
(a) $\beta_{e,y}=1.0$ m; (b) $\beta_{e,y}=1.5$ m; 
(c) $\beta_{e,y}=2.0$ m; (d) $\beta_{e,y}=3.0$ m; and 
(e) $\beta_{e,y}=4.0$ m.

Figure 4. Power spectrum of the centroid motion of the $e^+$ beam in (a) 
horizontal and (b) vertical direction for the case of HERA beam experiment
at $\xi_{e,y}=0.272$ ($\beta_{e,y}=4.0$ m). The beam centroid motion was 
calculated during the beam simulation. Note that in this case the incoherent
beam-beam tune shifts for the two IPs are 0.082 and 0.544 in the horizontal 
and vertical plane, respectively.

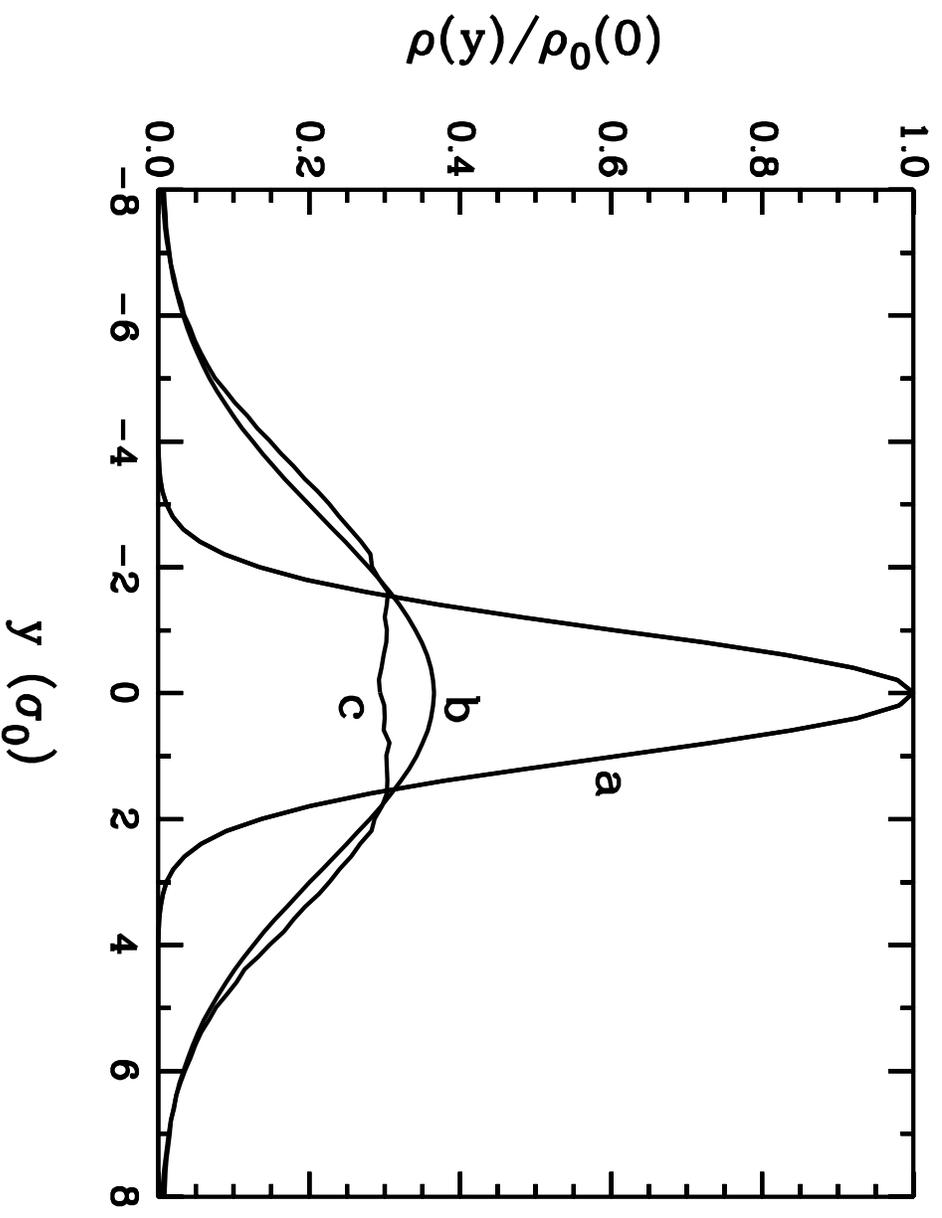
Figure 5. The projection of the particle distribution of the $e^+$ beam in 
the vertical direction obtained by the beam simulation for the case of HERA 
beam experiment at $\xi_{e,y}=0.272$ ($\beta_{e,y}=4.0$ m). (a) The initial
Gaussian distribution, (b) the distribution at the 5000th turn, and (c) the
Gaussian distribution of which the standard deviation is the same as that of
the beam distribution in (b).

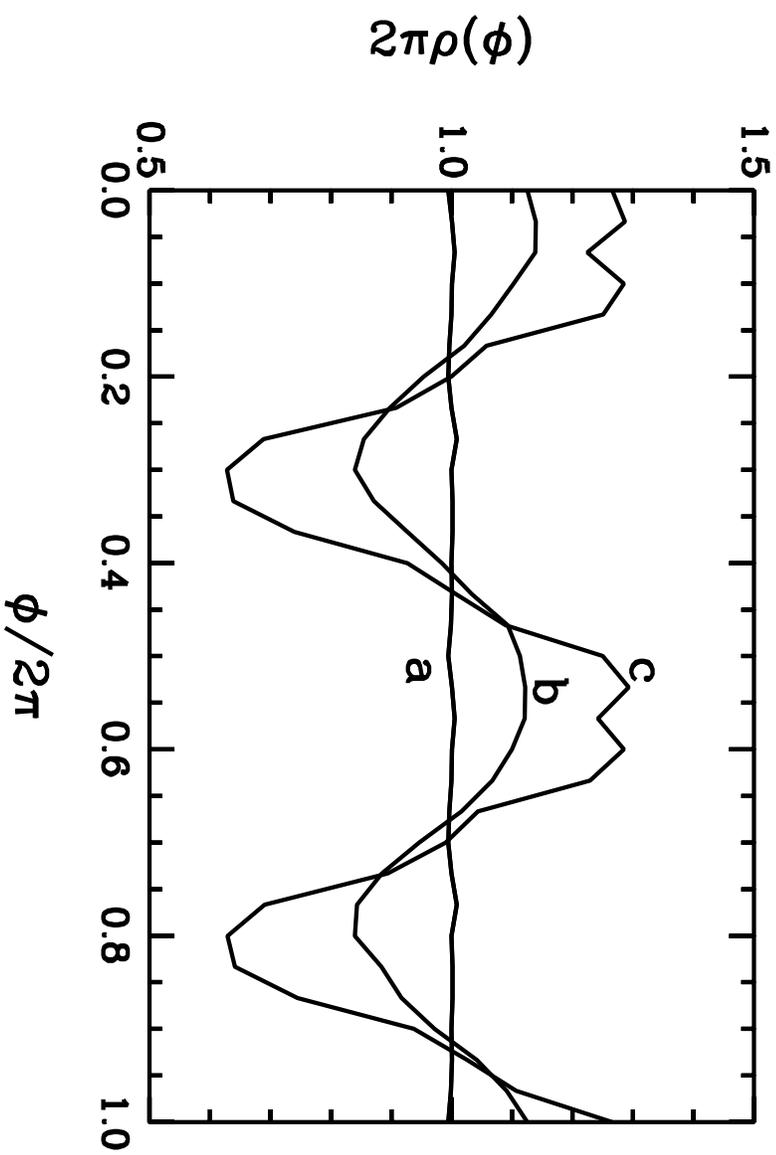
Figure 6. The angle dependence of the particle distribution of the $e^+$ beam
obtained by the beam simulation for the case of HERA beam experiment at 
$\xi_{e,y}=0.272$ ($\beta_{e,y}=4.0$ m). (a) The initial Gaussian distribution,
(b) the $\phi_x$-dependence and (c) the $\phi_y$-dependence of the distribution 
at the 5000th turn.

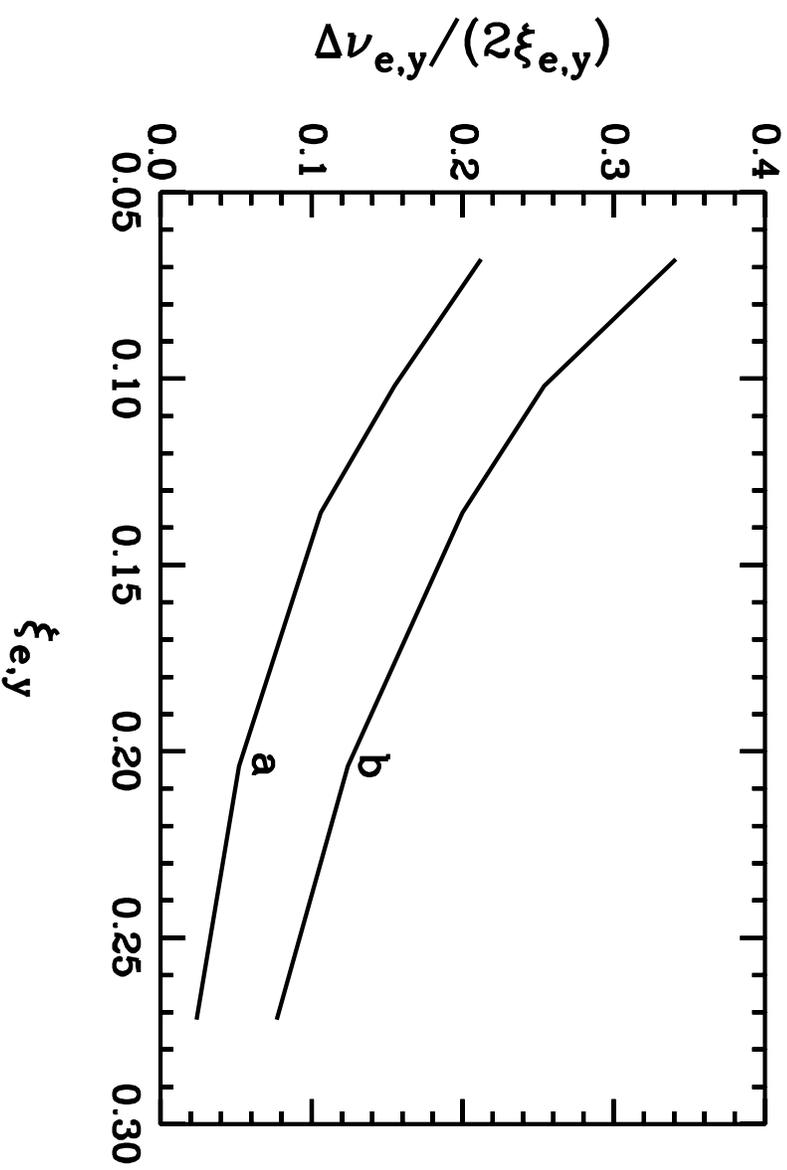
Figure 7. The ratio of the coherent and incoherent beam-beam tune shift as a 
function of the beam-beam parameter of the $e^+$ beam in the vertical plane 
calculated from (a) the beam-beam simulation and (b) the rigid-beam model 
with the Gaussian distribution. 

Figure 8. Eigenfrequencies of Eq. (B13) for the case of (a) 
$\nu_{1,x}=\nu_{2,x}$, $\xi_{1,x}=\xi_{2,x}$ and (b) the HERA beam 
experiment. In (a), the dashed line marks the lattice tune that corresponds to 
the $0$-mode. The single isolated line on the left is the coherent frequency 
that corresponds to a ratio of the coherent and incoherent beam-beam tune 
shift of 1.35. In (b), the dashed lines mark the lattice tunes of the $e^+$ 
beam (the left line) and $p$ beam (the right line), respectively. The single 
isolated line on the right is the degenerated eigenfrequencies for the $p$ 
beam and the band on the left is for the $e^+$ beam. The vertical axis has 
no physical meaning.

Figure 9. $|C_k|^2$ as a function of $\nu_k$ for the cases of Fig. 8. 
The arrows indicate the coherent frequencies.

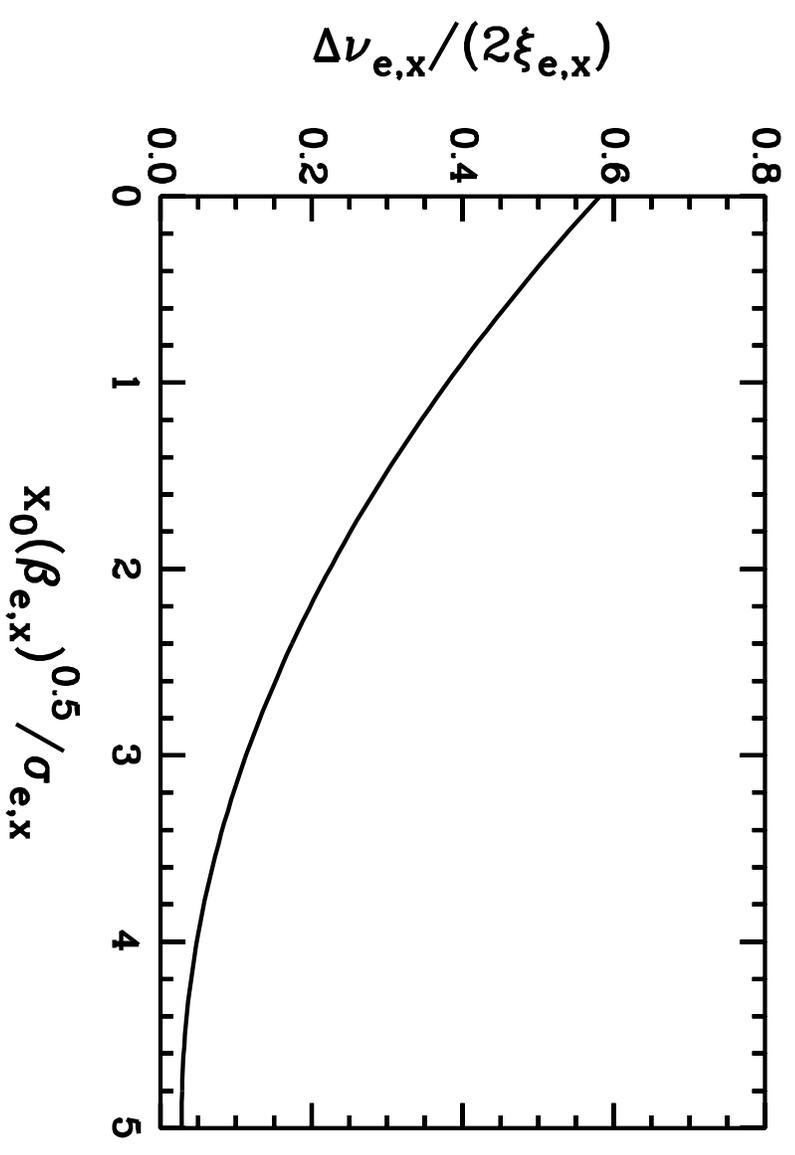
Figure 10. Calculated coherent beam-beam tune shift by using the linearized 
Vlasov equation as a function of initial kick $x_0$ on the distribution of 
the $e^+$ beam [see Eq. (B15)] for the case of the HERA experiment at 
$\beta_{e,y}=1.0$ m.

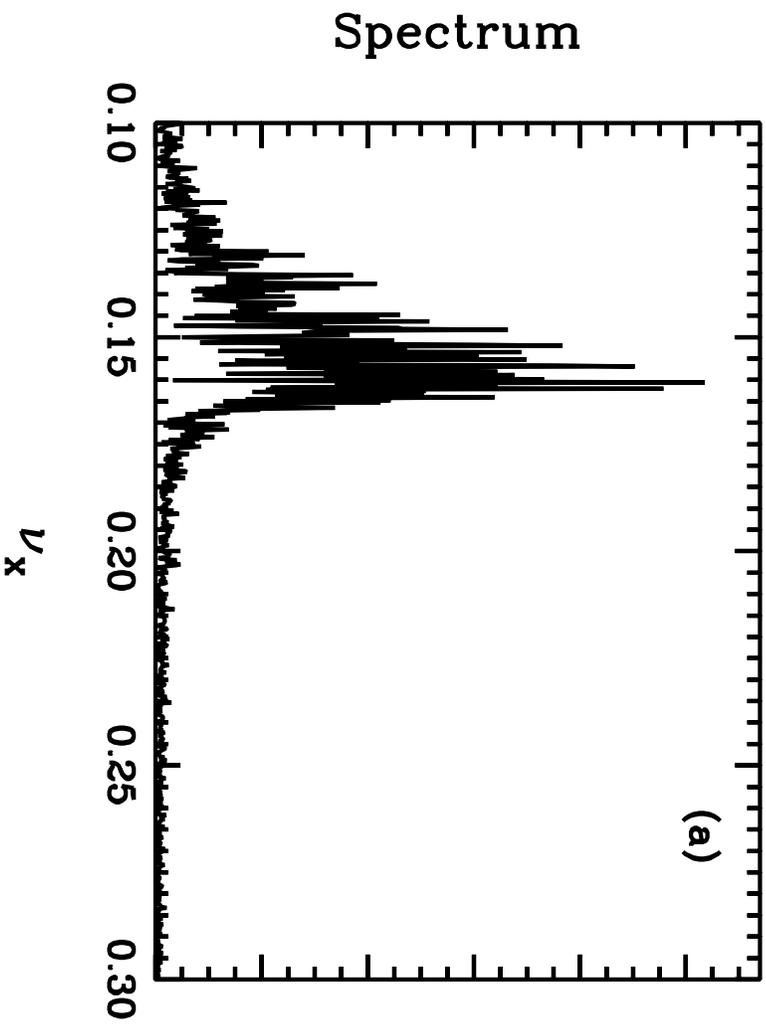
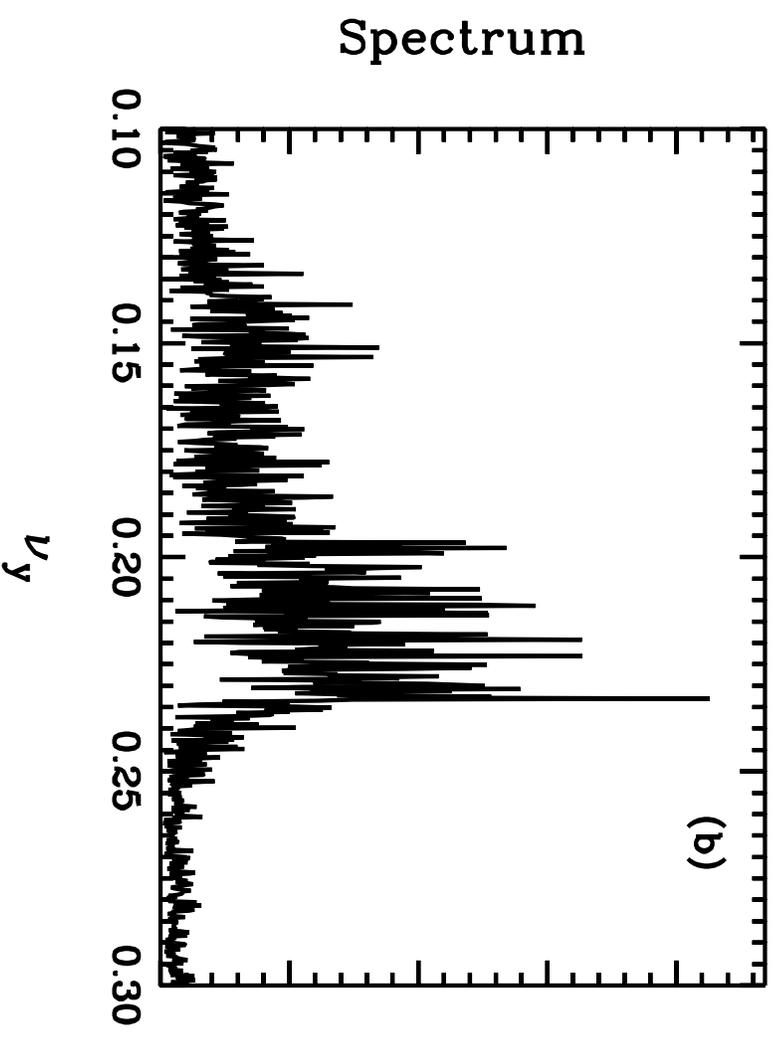
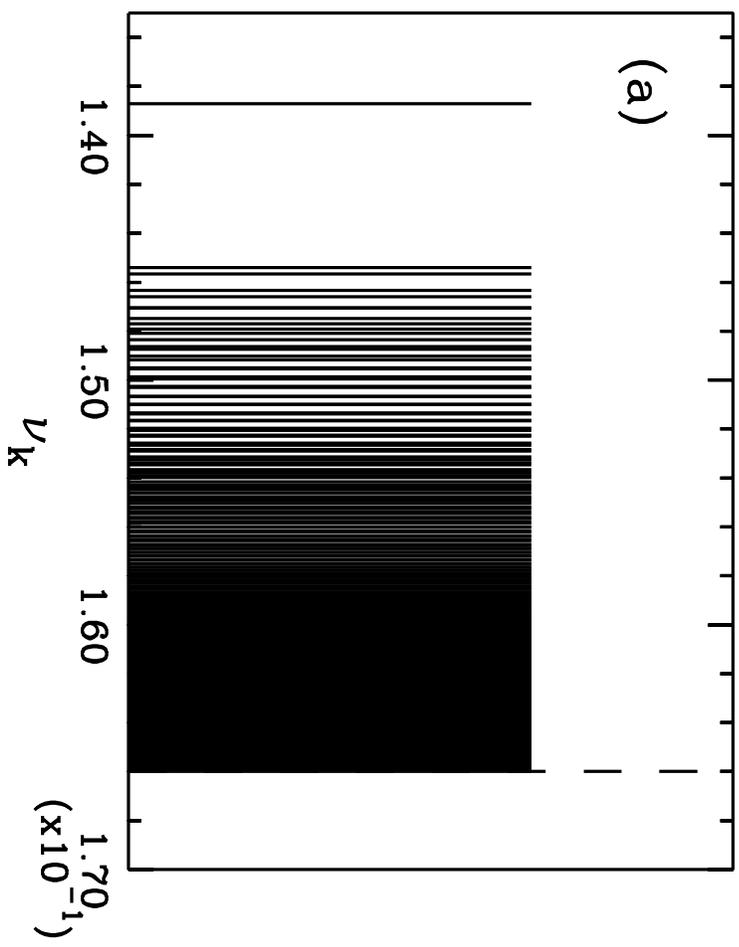
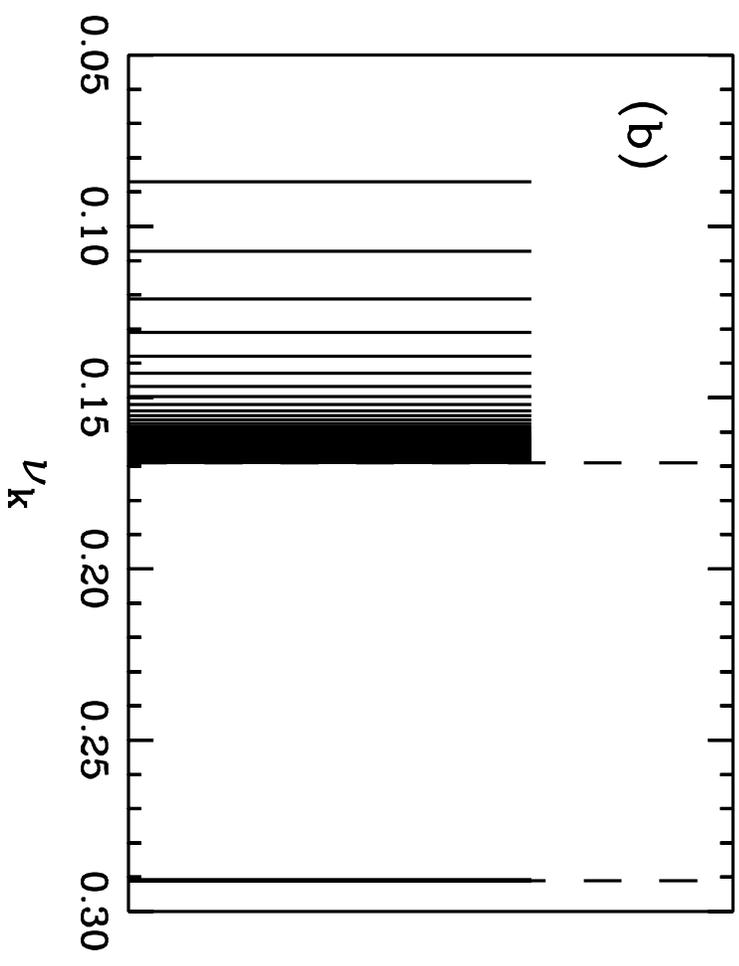
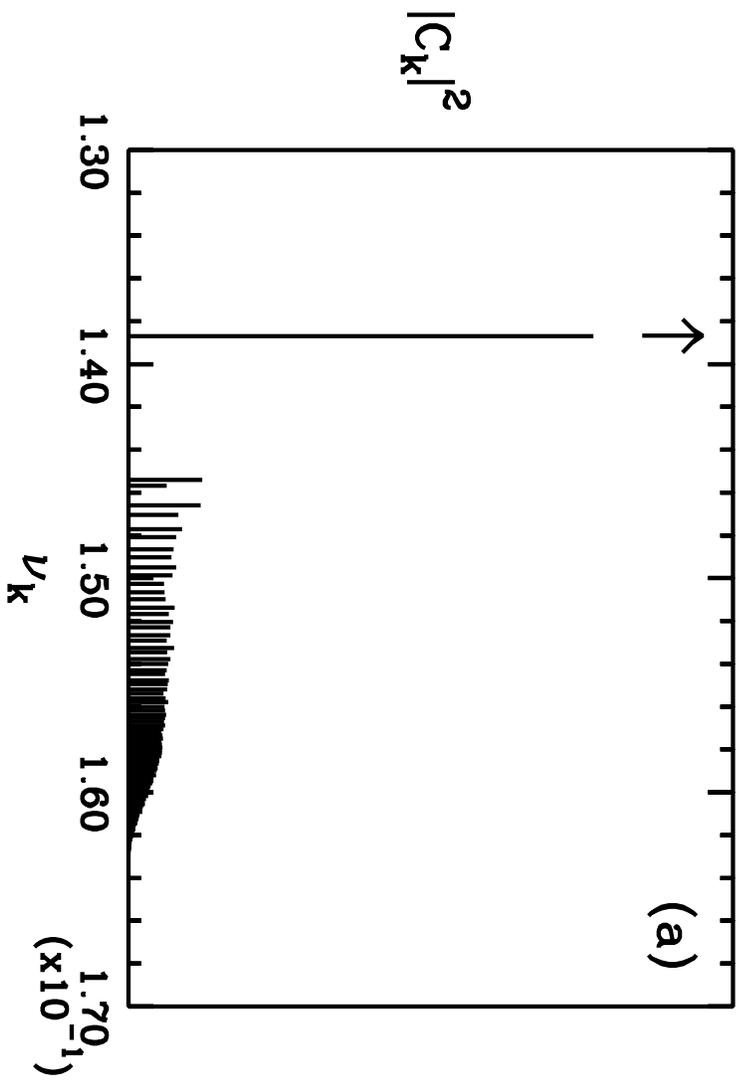
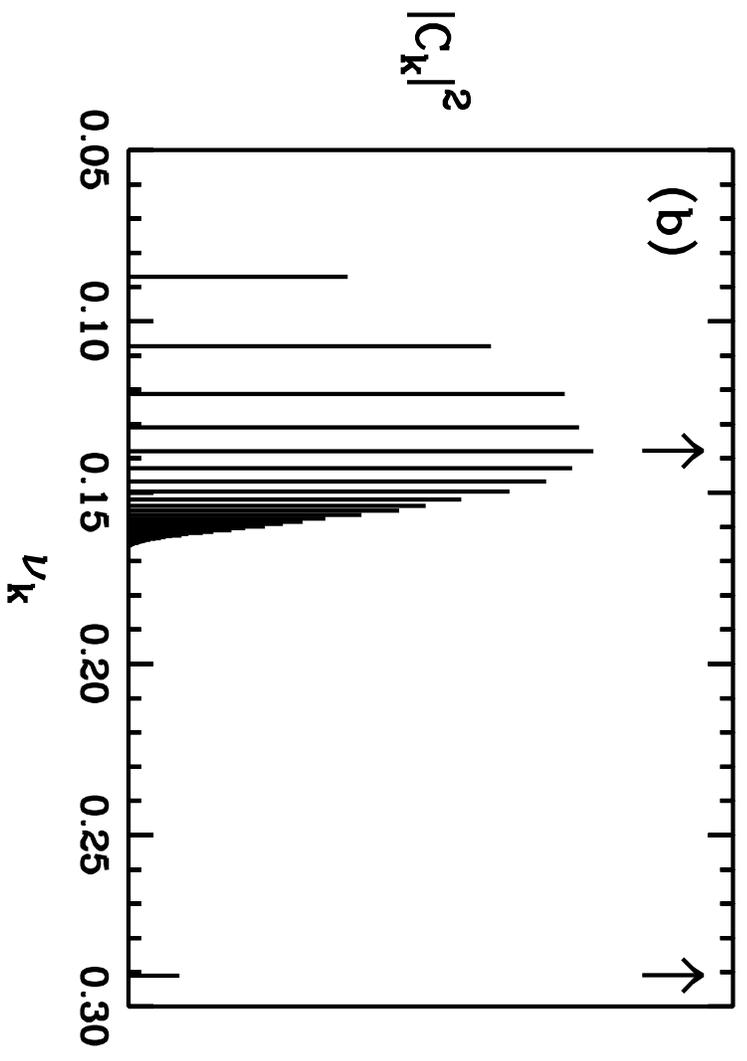
\end{document}